\documentclass[1p]{elsarticle}
 
\usepackage{graphicx}
\usepackage{amsmath}
\usepackage{amsfonts}
\usepackage{caption}
\usepackage{subcaption}
\usepackage{multirow}

\newcommand{\statevector}{\mathbf{q}}
\newcommand{\eigenmode}{\mathbf{p}}
\newcommand{\residue}{\mathbf{r}}

\newcommand{\extraction}{\mathcal{E}}
\newcommand{\permittivity}{\varepsilon}
\newcommand{\permeability}{\mu}
\newcommand{\Ecoeff}{{\mathbf{E}}}
\newcommand{\Hcoeff}{{\mathbf{H}}}

\journal{Journal of Computational Physics}

\begin{document}

\begin{frontmatter}

\title{Causal--Path Local Time--Stepping in the Discontinuous Galerkin Method for Maxwell's equations}

\author[ugr]{L. D. Angulo}
\author[eads]{J. Alvarez}
\author[osu]{F. Teixeira}
\author[ugr]{A. R. Bretones}
\author[ugr]{S. G. Garcia}

\address[ugr]{
 Department of Electromagnetism and Matter Physics,
 University of Granada,
 Granada 18071, 
 Spain
 (e-mail: lmdiazangulo@ugr.es)
}

\address[eads]{
 Cassidian, 
 EADS-CASA, 
 Getafe 28906, 
 Spain 
 (e-mail: jesus@ieee.org)
}

\address[osu]{
 ElectroScience Laboratory and the Department of Electrical and Computer Engineering,
 The Ohio State University,
 Columbus, Ohio, 43212 USA 
 (e-mail: teixeira@ece.osu.edu)
}

\begin{abstract}
We introduce a novel local time-stepping technique for marching-in-time algorithms.
The technique is denoted as Causal-Path Local Time-Stepping (CPLTS) and it is applied  for two time integration techniques: fourth order low--storage explicit Runge--Kutta (LSERK4) and second order Leapfrog (LF2). The CPLTS method is applied
 to evolve Maxwell's curl equations using a Discontinuous Galerkin (DG) scheme for the spatial discretization.

Numerical results for LF2 and LSERK4 are compared with analytical solutions and the Montseny's LF2 technique.
The results show that the CPLTS technique improves the dispersive and dissipative properties of LF2-LTS scheme. 

\end{abstract}


\end{frontmatter}

\section{Introduction}
 \label{s:introduction} 
 Many time--stepping algorithms have been proposed in order to improve the performance of Discontinuous Galerkin (DG) based schemes by increasing the maximum time step while preserving stability. There are usually two kinds of strategies used for this purpose: to use implicit schemes \cite{dolean10, piperno07} or, to use a explicit local time--stepping (LTS) technique \cite{alvarez12, pebernet08, dosopoulos10, piperno07, montseny08, diaz09, grote10, kaser07}. 
The advantage of LTS schemes versus implicit strategies is that the former can be used recursively and easily paralellized. This leads to an improvement in performance independent of the relative size distribution of the elements in the mesh. 
Additionally, time integration algorithms may have other constraints on the time--step arising from accuracy considerations and other inherent time scales such as in dispersive media \cite{sarmany13} or when hybridized with network/lumped elements models \cite{zhao12}, LTS techniques can also contribute to mitigate these problems in a simple and straightforward way.

When  a second order convergent spatial discretization is used, the most commonly used time integration method is the second--order leapfrog (LF2) algorithm. Several authors \cite{pebernet08, dosopoulos10} use a LF2-LTS scheme proposed by Montseny \cite{montseny08} consisting on using the last known values of the fields on the larger time stepped region each time that the smaller one needs a field value. Piperno \cite{piperno07} adopts a similar approach based on a Verlet scheme. 
Alvarez \cite{alvarez12, alvarez13, alvarez13b, alvarez13c, alvarez13d} contributed with a novel approach to perform LTS in LF2 schemes whereby an interpolation between the fields is used in an interface between the larger and smaller time--stepped regions.  A rigorous demonstration of the stability and dispersive properties of these schemes is still an open problem. 

Diaz and Grote \cite{diaz09, grote10} implemented a rigorous study on the stability and dispersion of LF-LTS high--order schemes applied to the second--order wave equation by means of an eigenvalue analysis. They found that the LTS introduces numerical dispersion and can produce instabilities if the global time step is not slightly reduced with respect to a classic implementation. The authors also found that the global stability could be improved by enlarging the smaller time--stepped region.
 
 For higher order methods, explicit Runge-Kutta (RK) algorithms \cite{hesthaven07, sarmany07, alvarez10, angulo11, niegemann09, diehl10} seem to be preferred with respect to LF schemes \cite{fahs09b}. Despite of their popularity, there are less works in the literature related to RK-LTS than to LF-LTS. For Maxwell's equations RK-LTS algorithms usually rely on interpolations at the interfaces using previously computed solutions \cite{hesthaven07} or arbitrary high-order derivatives (ADER) schemes \cite{kaser07, taube09, liu10}. 
 
 In this paper we present a novel LTS technique that can be applied to a large variety of time integration algorithms. It does not need interpolation between computed solutions and nor directly uses any previously known values. Numerical results showing comparisons with analytical solutions for applications on a second--order Leap-Frog (LF2) and on a fourth--order Low Storage Explicit Runge--Kutta scheme (LSERK4) are shown to demonstrate the advantages of the proposed LTS technique. 
 
 \section{Discontinuous Galerkin Semidiscretization}
  \label{sec:DGSemidiscretization}
  Maxwell's curl equations for source--less homogeneous media can be written as
  \begin{align}
   \label{eq:MaxwellEquations}
   \vec{\nabla} \times \vec{E} & = - \permeability \partial_t \vec{H} \nonumber \\
   \vec{\nabla} \times \vec{H} & = \permittivity \partial_t \vec{E}
  \end{align}
 For simplicity, in our discussion we will assume that $\permittivity$ and $\permeability$ do not vary in the computational domain, and use a system of units where $\permittivity = \permeability = 1$.
 
 We tessellate the computational domain with $k = 1, \hdots, K$ non--overlapping tetrahedrons. In each of those, we apply the Discontinuous Galerkin's formalism \cite{hesthaven07,alvarez10, angulo11} to obtain
  \begin{align}
   \label{eq:MaxwellGalerkin}
   \mathcal{M}_k \partial_t \Ecoeff_k(t) & 
    +  \mathcal{S}_k \Hcoeff_k(t) 
    - \sum_{f}  \mathcal{F}_{kf} \Hcoeff_{kf}^*(t)
  = 0 \nonumber \\
   \mathcal{M}_k \partial_t \Hcoeff_k(t) & +  \mathcal{S}_k \Ecoeff_k(t)
    - \sum_{f}  \mathcal{F}_{kf} \Ecoeff_{kf}^*(t)  = 0
 \end{align}
  With $\mathcal{M}$ being the mass matrix, $\mathcal{S}$ the spatial semidiscretization of the curl operator and $\mathcal{F}_f$ the lift operator for face $f$. $\Ecoeff$ and $\Hcoeff$ are column vectors containing all the degrees of freedom for the electric and magnetic field respectively. $\Ecoeff^*$ and $\Hcoeff^*$ are the numerical fluxes.
  
  We define a state vector $\statevector_k = [\Ecoeff_k \ \Hcoeff_k]^T$ containing all the $N_k$ degrees of freedom of element $k$. With this definition, we can rewrite system (\ref{eq:MaxwellGalerkin}) as a single equation that governs the time evolution of the system,
  \begin{equation}
   \label{eq:evolution}
   \partial_t \statevector_k(t) = 
   - (\mathcal{M}^q_k)^{-1} \left({\mathcal{S}}^q_k \statevector_k(t)
    - \sum_{f} \mathcal{F}^q_{kf}  
    \left( \extraction_{kf} \statevector_k(t) - \extraction_{kf_+} \statevector_{kf_+}(t) \right)
    \right)
  \end{equation}
  The DG method gives us some freedom in the selection of the operators  $\extraction_{kf}$ and $\extraction_{kf_+}$ as long as it respects the properties of consistency, continuity, and monotonicity needed for the numerical flux \cite{shu10}. If this operator is block diagonal with all its components being $1/2$, we will say that the semi-discrete scheme is using a centered flux and therefore is numerically non-dissipative \cite{sherwin99, alvarez12}. On the other hand, if these operators are non-block diagonal we will say that the flux is being penalized and therefore the semi-discrete scheme is numerically dissipative. We will mostly focus on a particular case of penalized flux: the upwind flux \cite{alvarez10, alvarez12d}, coming from the solution of the Riemann problem.
   
  When using penalized fluxes some dissipation is introduced and more operations are needed to compute the flux terms. However, introducing such penalization is known to improve numerical dispersion and suppress spurious modes \cite{montseny08, alvarez10, alvarez12, hesthaven07, hesthaven04}. Thus, the penalized fluxes approach is usually preferred for simulations not requiring very long integration times.  
  
  To simplify the discussion further we will change the basis of the vector space using an invertible operator $\mathcal{P}_k$ on equation (\ref{eq:evolution}) that diagonalizes only the locally applied operators,
  \begin{equation}
   \label{eq:diagionalization}
   \mathcal{W}_k = - \mathcal{P}_k^{-1} (\mathcal{M}^q_k)^{-1} ( \mathcal{S}^q_k 
    - \sum_{f}  \mathcal{F}^q_{kf}  
  \extraction_{kf}  ) \mathcal{P}_k
 \end{equation}
 We can also define the eigenmodes as
 \begin{equation}
   \eigenmode_k = \mathcal{P}^{-1}_k \statevector_k
 \end{equation}
 and the external operators as
 \begin{equation}
  \mathcal{V}_{kf} =  - \mathcal{P}_k^{-1} (\mathcal{M}^q)_k^{-1} 
       \mathcal{F}^q_{kf}  
     \extraction_{kf_+} \mathcal{P}_k
 \end{equation}
 This change of basis let us write equation (\ref{eq:evolution}) in the following compact form
 \begin{equation}
  \label{eq:diagevolution}
  \partial_t \eigenmode_k(t) = \mathcal{W}_k \eigenmode_k(t) + \sum_f \mathcal{V}_{kf} \eigenmode_{kf_+}(t)
 \end{equation}

 \section{Time integration}
 In the following discussion, we will focus on two time integration methods that are also the most popular choices in conjunction with DG semidiscretizations.
 
 \subsection{Second-order Leap-Frog (LF2)}
  The second-order leap-frog method \cite{dosopoulos12} is applied by alternately evolving the $\Ecoeff^n$ and $\Hcoeff^{n+1/2}$ fields, arbitrarily defined at times $t_n$ and $t_n + \Delta t/2$ respectively. This implies that we do not have a fully defined state vector in the sense of eq. (\ref{eq:evolution}) for a given time $t$. To obtain the future values from a present state the following algorithm is applied
  \begin{align}
   \Ecoeff^{n+1} & = \Ecoeff^n + \Delta t \ \mathcal{L}_h \left(\Hcoeff^{n+1/2}, \Ecoeff^n \right) \nonumber \\
   \Hcoeff^{n+3/2} & = \Hcoeff^{n+1/2} + \Delta t \ \mathcal{L}_h \left(\Ecoeff^{n}, \Hcoeff^{n+1/2} \right)
  \end{align}
  With $\mathcal{L}_h$ being a function representing the result of applying the spatial semi--discretization.
  When centered fluxes are used, the operator $\mathcal{L}_h$ only uses $\Hcoeff^{n+1/2}$ or $\Ecoeff^{n}$ as arguments. This implies that the scheme is reversible in time and will preserve energy as long as the time step used  is below a maximum value $\Delta t_k$ set by a CFL-like condition \cite{dolean10, piperno07, dosopoulos12}.
  
 \subsection{Low-Storage Explicit Runge--Kutta (LSERK4)}
 The second method that we will use in our discussion is the five-stage fourth-order Explicit Runge-Kutta method (LSERK4) \cite{hesthaven07, diehl10}. This method states that for a given vector representing the state of the system, i.e. $\eigenmode_k(t) = \eigenmode^n_k$ we can find an approximate solution state $\eigenmode_k(t + \Delta t) = \eigenmode_k^{n+1}$ applying the following algorithm
 \begin{align}
 \label{eq:LSERKScheme}
  \eigenmode_k^{(0)} & = \eigenmode_k^n, \nonumber \\
  \residue^{(i)} & = a_i \residue^{(i-1)} 
  + \Delta t \left( \mathcal{W}_k \eigenmode_k^{(i-1)} + \sum_f \mathcal{V}_{kf} \eigenmode_{kf_+}^{(i-1)} \right) \nonumber, \\
  \eigenmode_k^{(i)} & = \eigenmode_k^{(i-1)} + b_i \residue^{(i)}, \nonumber \\
  \eigenmode_k^{(n+1)} & = \eigenmode_k^{(5)}
 \end{align}
 with $i \in [1, ..., 5]$ and the coefficients $a_i$, $b_i$ and $c_i$ taking the values indicated in Table \ref{table:rkConstants}.
 The LSERK4 scheme is one of the most used methods in high--order Discontinuous Galerkin semi--discretizations, because it introduces low dispersion and dissipation. Contrary to other RK implementations, the low--storage version requires the storage of only two times the number of degrees of freedom in the scheme at the expense of one additional stage. RK methods are constrained by the spectra of the operator $\mathcal{W}_k$, i.e. all the eigenvalues of $\mathcal{W}_k$ must lie inside of the stability region of the RK scheme. Consequently, the time step must be chosen sufficiently small, e.g. for a nodal basis the following inequality must hold \cite{hesthaven07}
 \begin{equation}
  \label{eq:maxTimeStep}
  \Delta t_k \leq \frac{C}{c_k} \min_{i} \frac{\Delta r_{ki}}{2}
 \end{equation}
 where $\min_{i} \Delta r_{ki}$ indicates the minimum distance between nodes in element $k$ and $c_k$ is the maximum speed of light in the element $k$.
 
Despite its many advantages, LSERK4 has a high computational cost and the numerical dissipation it introduces can be a factor depending on the application.

\begin{table}[h]
 \centering
 \begin{tabular}{c c c c}
  \hline
  $i$ & $a_i$ & $b_i$ & $c_i$\\
  \hline 
  \\
  1 & 0  & $\frac{1432997174477}{9575080441755}$ &  0 \\
  \\
  2 & - $\frac{567301805773}{1357537059087}$ 
           & $\frac{5161836677717}{13612068292357}$
                & $\frac{13612068292357}{9575080441755}$ \\
                \\
  3 & - $\frac{2404267990393}{2016746695238}$ 
           & $\frac{1720146321549}{2090206949498}$ &
              $\frac{22526269341429}{6820363962896}$ \\
              \\
  4 & - $\frac{-3550918686646}{2091501179385}$ 
           & $\frac{3134564353537}{4481467310338}$
              & $\frac{2006345519317}{3224310063776}$ \\
              \\
  5 & - $\frac{1275806237668}{842570457699}$ 
          & $\frac{2277821191437}{14882151754819}$
              & $\frac{28032321613138}{2924317926251}$ \\
              \\
  \hline
 \end{tabular}
 \caption{Coefficients for the low-storage five-stage fourth-order Explicit Runge--Kutta method (LSERK4) \label{table:rkConstants}}
\end{table}

 \section{The Causal--Path LTS technique} 
 \label{sec:CausalPath}
 In this section we introduce the Causal--Path technique as a novel way of performing LTS in different time integration techniques. We require two basic properties for the time integration technique: 

 \begin{enumerate}
  \item It has to provide a fully defined state $\statevector_k(t)$ for each element.
  \item The next state $\statevector_k(t+\Delta t)$ can be explicitly computed from a neighbourhood of elements.
 \end{enumerate}
 
 As a first step we will organize the elements in different groups, called tiers, according to their time steps denoted as $\Delta t^m$. An element $k$ will belong to a tier $m = [0, \hdots, N_m - 1]$ if its maximum time step $\Delta t_k$ is such that 
 
 \begin{equation}
  \label{eq:tierAssorting}
  \Delta t^m \leq \Delta t_{k} < \Delta t^{m+1}
 \end{equation}
 
 In order to compute the next time step, we need to use the field values at local and neighbor elements, $\eigenmode^{(m,i-1)}_k$ and $\eigenmode^{(m,i-1)}_{kf_+}$. If there is no connection with other elements belonging to a lower tier, we can evolve all the elements in $m$ using their $\Delta t^m$. However, in the border between a tier $m$ and $m+1$ we can not apply the direct algorithm because the value $\eigenmode^{(m,i-1)}_{k^mf_+} = \eigenmode^{(m,i-1)}_{k^{m+1}}$ has not been computed.

  The strategy that we propose is to compute the values $\eigenmode^{(m,i-1)}_{k^{m+1}}$ using $\Delta t^{m-1} = h_{i-1} \Delta t^{m}$ as time step wherever they are necessary. If to do that, we need additional neighbour values that have not been computed, and we recursively apply this idea until a known value is found.  Thus, starting from $m=0$ we can compute all the stages needed to evolve it before starting with the tier $m=1$ and so on. Finally, the values $\eigenmode^{(m,i-1)}_{k^{m+1}}$ are casted aside and the upper tiers uses the original values from the lower tier.

\begin{figure}[h]
  \centering
  \includegraphics[width=0.55\textwidth]{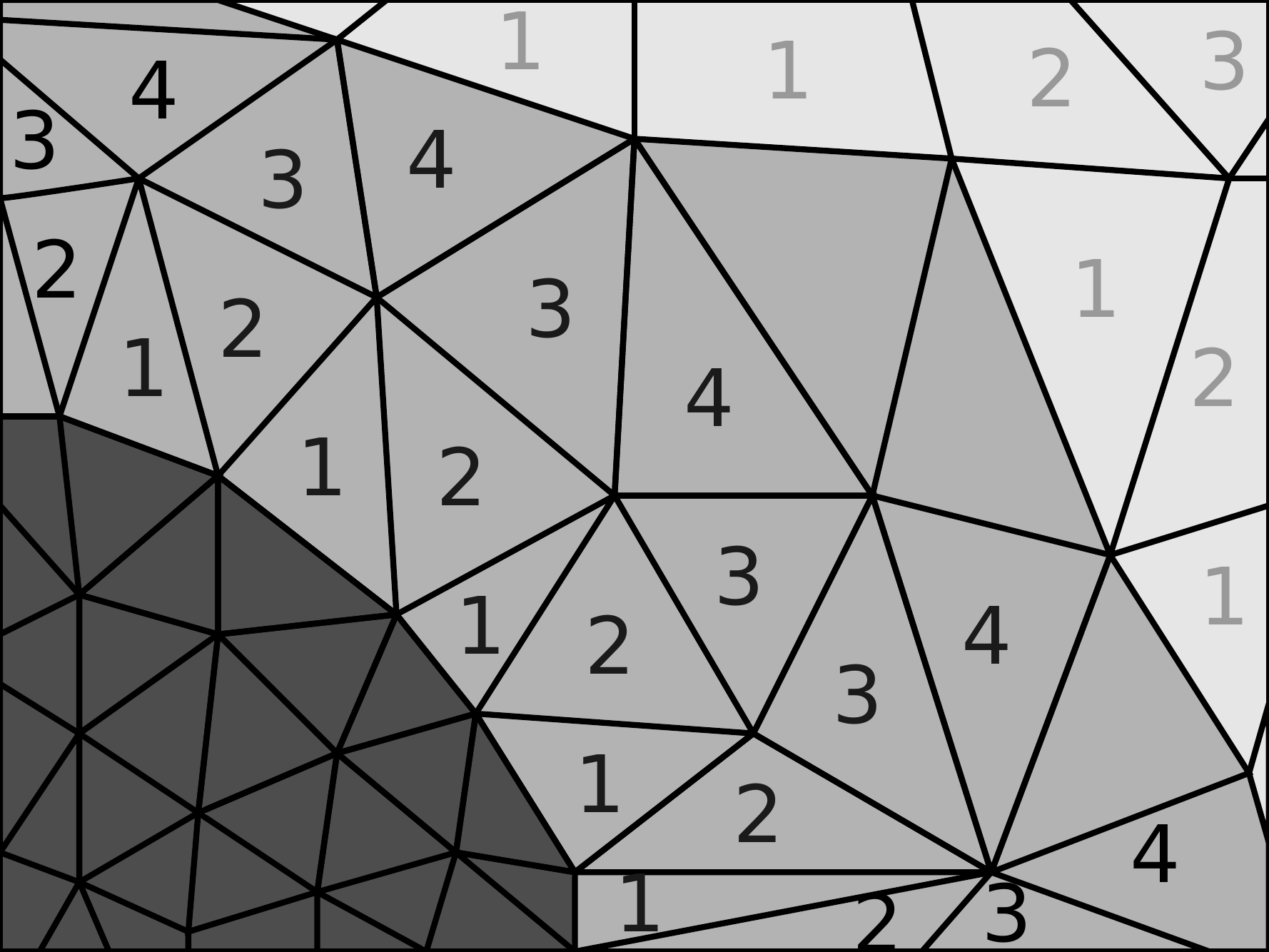}
  \caption{This figure illustrates the concept of 4--depth neigbourhood of two different regions. The darker colors indicate elements belonging to a lower temporal tier and thus have a smaller time step.}
  \label{fig:meshneighbourhood}
 \end{figure}

  To compute the next time step values in each of the $N_m$ tiers we may need to compute $N_s$ stages in all the elements of tier $m$. We will also need to compute intermediate stages between the stages in the $m+1$ tier. So, in order to avoid a possible interleaving with other higher tiers, we impose that the $(N_s-1)$--depth neighbourhood of a tier $m$ is only composed of elements belonging to tier $m+1$ or $m-1$. This additional condition for the tier assortment is illustrated in Figure \ref{fig:meshneighbourhood}.
     
 The implementation of this algorithm may seem difficult at a first glance; however, the recursive nature of the algorithm allows us to make use of recursive calls to the function used to evolve the system. Every time the function is called, we pass the information about the tier in which this is being computed and the time step that has to be applied. So starting from a call to evolve the $N_m$ tier for a given time step $\Delta t$, the function will recursively call itself on each of the stages of the algorithm passing $N_m-1$ and $h_i \Delta t$ as arguments and evolving its corresponding tier elements. This technique also requires that the degrees of freedom in the region being interfaced are saved in the higher tier.
Note that no interpolation of field values is necessary and only past field values generated by the discretization itself are utilized.

  In the next sections we describe two examples of the CPLTS technique, applied to the LF2 and LSERK4 algorithms, together with illustrations to clarify the concepts.
 
 \subsection{LF2-CPLTS}
 \label{ssec:LF2-CPLTS}
\begin{figure}[h]
  \centering
  \includegraphics[width=0.75\textwidth]{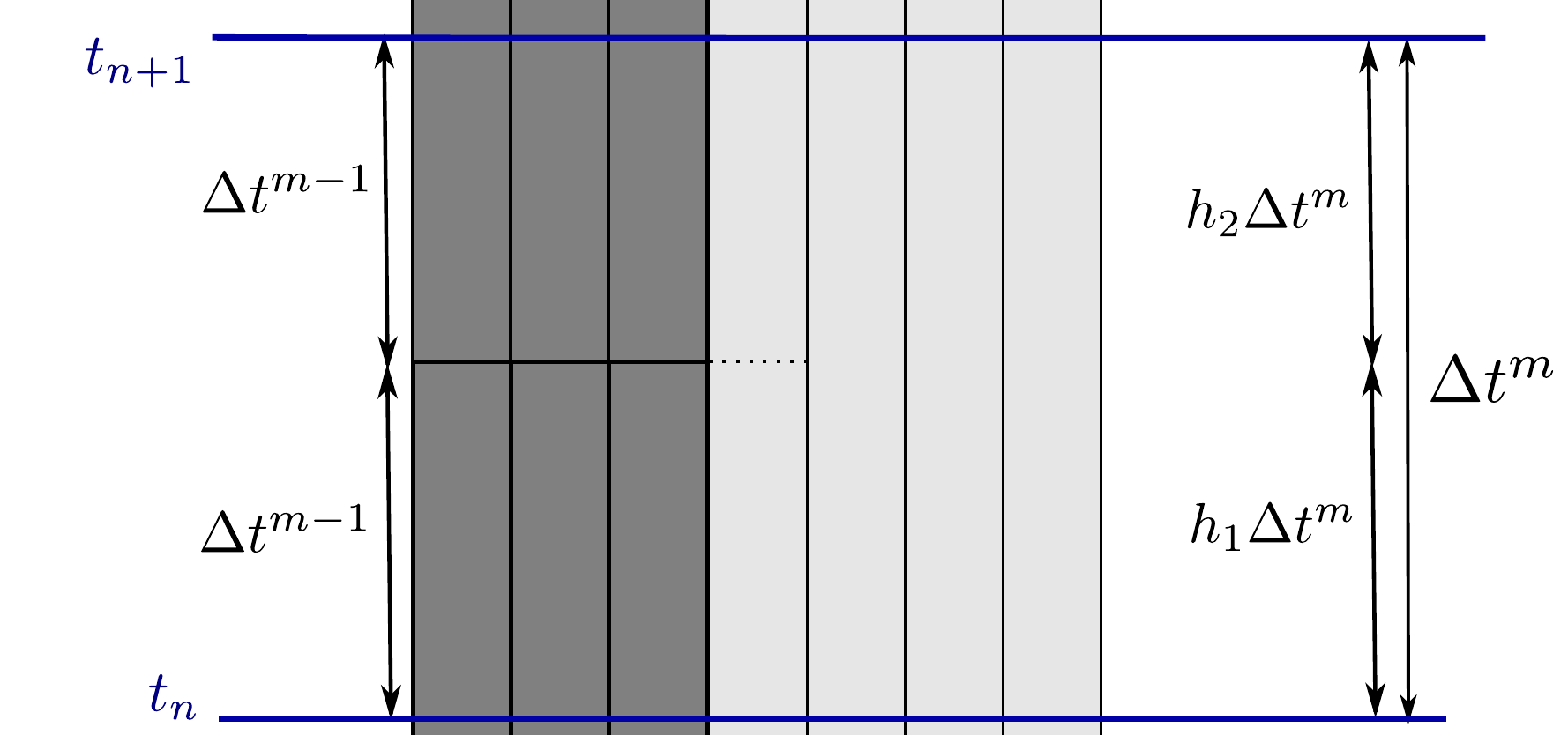}
  \caption{Schematic view of the LF2-CPLTS algorithm particularized to the case $h_1 = h_2 = 1/2$. Vertical lines indicate boundaries between elements. Dashed horizontal lines are intermediate stages. Continuous horizontal lines indicate time steps.}
  \label{fig:lf-cplts}
 \end{figure}

  Since the LF2 performs iterations using a single stage we can create any distribution of $N_s$ intermediate stages in the higher tiers to fit the evaluations needed by the smaller tiers. The time--steps of the intermediate stages would then be $ h_i \Delta t^{m+1,i} =  \Delta t^m $, with $h_i > 0$ and the restriction $\sum_i^{N_s} h_i = 1$. The choice of $h_i = 1 / N_s$ would be the most favourable in terms of computational cost. Figure \ref{fig:lf-cplts} shows an schematic view of this scheme applied to the case $h_1 = h_2 = 1/2$.
  Note that this freedom in choosing $h_i$ is an improvement compared with the Montseny's scheme \cite{montseny08}, which is constrained by the condition $\Delta t^m = \Delta t^{m-1}(1 + 2k)$. This is also an improvement with respect to the Verlet--Piperno's scheme \cite{piperno07} in which $\Delta t^{m+1} = 2 \Delta t^m$, and it allows our scheme to adapt to the different transitions as necessary; however, for the sake of simplicity we will not consider these cases here.
  
 On the other hand, we need both values of $\Ecoeff$ and $\Hcoeff$ at same time instants in order to find a fully--defined state of the system at any given stage $\eigenmode^{(m,i)}$.
  In other words, we can't apply this LTS technique computing only $\Ecoeff(t_n)$ and $\Hcoeff(t_n +\Delta t /2 )$ because to compute the intermediate value of a lower tier, let's say $\Ecoeff^{m-1}(t_n + \Delta t^m/2 )$ we would need the values of the magnetic field $\Hcoeff^m(t_n)$. To overcome this issue we need to apply LF2 twice, doubling the computational costs with respect to the conventional approach.
 
 When we apply this scheme to a non--dissipative semi--discretization (e.g. DG with centered flux) we find that the scheme is unstable showing growing high--frequency numerical modes.
 
  The introduction of a penalized flux solves this problem through higher frequency damping \cite{sarmany07, sherwin99}.
 
 \subsection{LSERK4-CPLTS}
 \label{ssec:LSERK4-CPLTS}
\begin{figure}[h]
  \centering
  \includegraphics[width=0.75\textwidth]{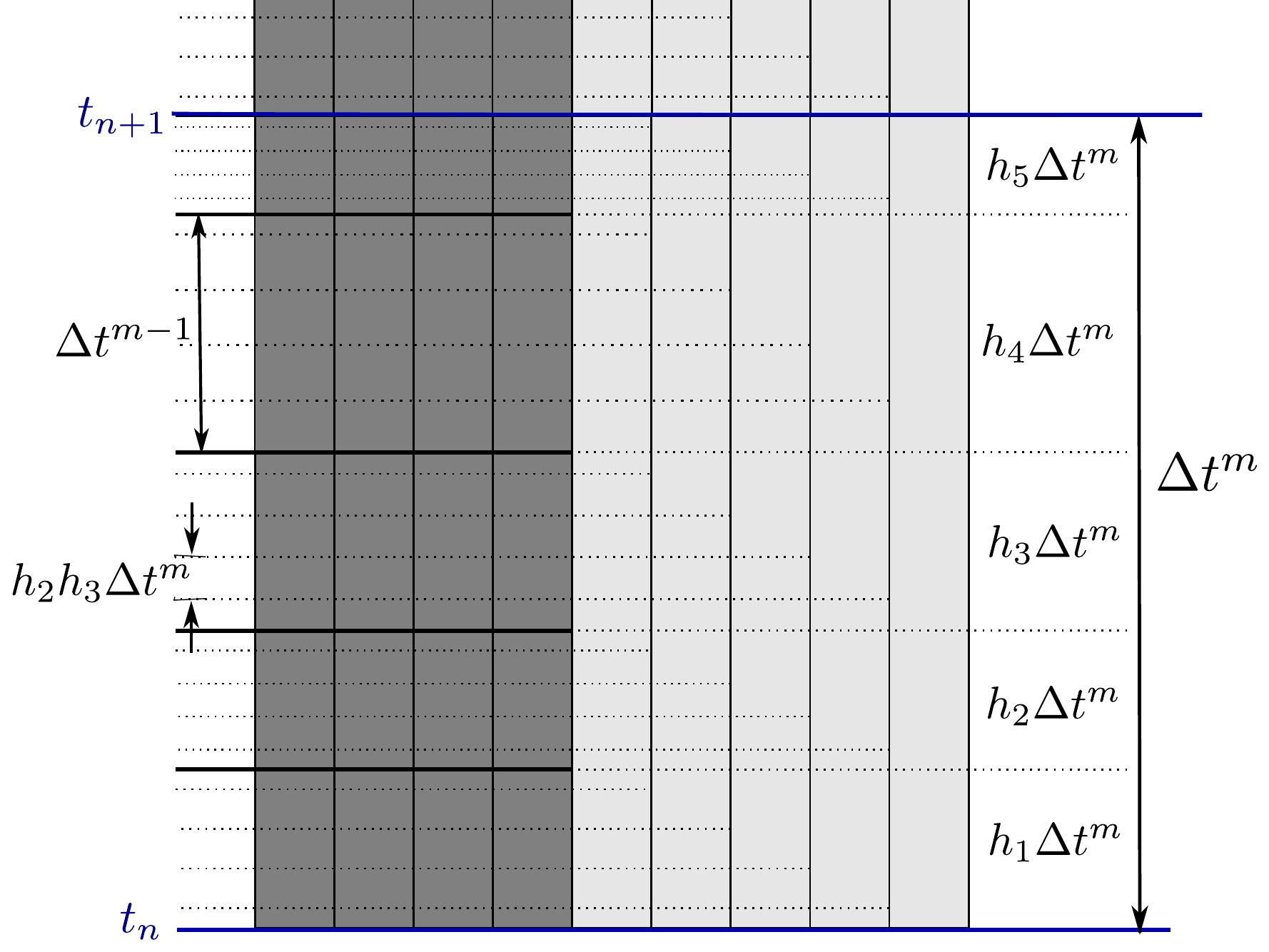}
  \caption{LSERK4-CPLTS sketch. Additional operations are made only in the 4-depth neighbourhood of the smaller tier region (darker). In this sketch $\Delta t^{m-1} = h_4 \Delta t^m$.}
  \label{fig:rk-lts}
 \end{figure}

 When the CPLTS technique is applied to an LSERK4 we note that the stages are not evenly distributed in time. As a result, we apply a variable time step in the lower tiers (Figure \ref{fig:rk-lts}). The values for $\Delta t^m$ and $\Delta t^{m-1}$ are chosen such that equation (\ref{eq:maxTimeStep}) is always enforced and therefore
 \begin{equation}
  \label{eq:tierSize}
  \max_i (h_i) \Delta t^m = \Delta t^{m-1} 
 \end{equation}
 with $\max_i (h_i)$ being the maximum stage size (for LSERK4 $\max_i (h_i) = h_4 = c_5 - c_4 = 0.336026 \simeq 1/3$). Whenever we compute intermediate stages in higher tiers we satisfy this condition because in higher tiers this condition is less restrictive. However, every time we apply this division, $N_s$ times more computational operations are needed to get a speed-up of about three times in the higher tier region. So, if the largest tier region is not at least $5/3$ times larger than the smallest we won't see any appreciable global speed-up.
 
 For this reason it seems preferable to organize the time tiers with $\Delta t^{m-1} = \Delta t^m / N_s$ rather than with the maximum stage size criteria.  
 By doing this, we are computing an stage in the lower tier region with a time-step bigger than is strictly allowed based on a conventional CFL-like criterion for the associated direct algorithm, which could be a source of potential instability.
On the other hand, the smaller stages in the lower tier compute the solution using a time-step smaller than the maximum allowed and thus introducing an additional numerical dissipation.
 We may then wonder if the additional dissipation introduced by the smaller stages offsets the potential for instability introduced by the larger.
 Note that as long as these effects are mostly kept limited to high frequency components (which are under-resolved anyway) the solution accuracy should not be impacted.
  In the next sections we perform some tests to assess the practical validity of this approach.

 \section{Numerical Results}
 In this section we present comparisons between results using the proosed CPLTS technique, the LF2-LTS technique introduced by Montseny \cite{montseny08}, classical implementations of the algorithms, and analytical solutions.
 
 For all cases we use nodal basis of order $P=2$ and numerical upwind fluxes as described in \cite{hesthaven07, alvarez10, angulo11, otin10}. This implies that we are using 60 degrees of freedom per element.
 The implementation has been performed with an in--house C++ code\footnote{Compiled with GNU C++ v4.6.3 using -O3 -ffast-math flags} with OpenMP parallelization\footnote{For more information visit: {http://www.ugrfdtd.es}}. GiD was used to obtain meshes and for pre and post-processing\footnote{For more information visit: {http://www.gidhome.com}}.
 Simulations for the reflection and resonance problems were performed using a single processor laptop with Intel(R) Core(TM)2 Duo CPU T9400  @ 2.53GHz processor and running Ubuntu 12.04 LTS. The RCS problem were run in a desktop computer with an Intel(R) Core(TM) i7-3960X CPU @ 3.30GHz processor with 12 cores and Ubuntu 10.04 LTS.

 \subsection{Reflection caused by a non-homogeneous mesh}
\begin{figure}
   \begin{subfigure}[b]{1\textwidth}
    \centering
    \includegraphics[width=\textwidth]{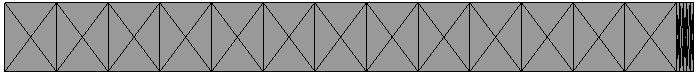}
    \caption{Single Interface 15:1 ratio.}
   \label{fig:sI-r15}
  \end{subfigure}    

  \begin{subfigure}[b]{\textwidth}
   \centering
   \includegraphics[width=\textwidth]{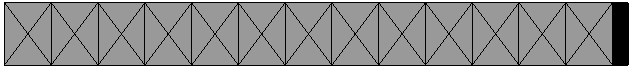}
   \caption{Single Interface 75:1 ratio.}
   \label{fig:sI-r75}
  \end{subfigure}
  
  \begin{subfigure}[b]{\textwidth}
   \centering
   \includegraphics[width=\textwidth]{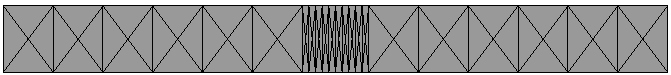}
   \caption{Slab 7.5:1 ratio.}
   \label{fig:slab-r7c5}
  \end{subfigure}

  \begin{subfigure}[b]{\textwidth}
   \centering
   \includegraphics[width=0.5\textwidth]{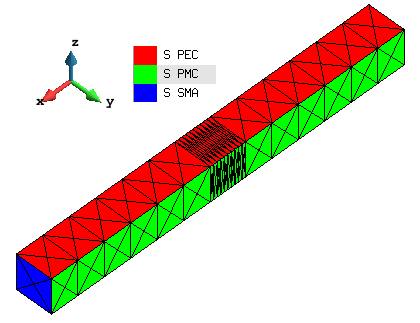}
   \caption{Boundary conditions.}
   \label{fig:slab-bc}
  \end{subfigure}
  
  \caption{Meshes used for the study numerical reflections by an inhomogeneous mesh.}
  \label{fig:mesh-for-reflection}
 \end{figure}

 The first example we present is an study of the numerical reflection caused by differences in the mesh size, a similar type of analysis can be found in \cite{chilton08, donderici08}.
 This type of analysis is important for LTS because it quantifies a source of additive noise on the results.
  Figure \ref{fig:mesh-for-reflection} shows the meshes used, together with an isometric view of the boundary conditions employed. A plane wave excitation with $z$-polarization is introduced in one of the ends of the computational domain and the other end is backed by an Silver-Mueller absorbing (SMA) boundary condition. The side--walls of the domain are Perfect Electric Conducting (PEC) and Perfect Magnetic Conducting (PMC) boundary conditions at the $xy$ and $xz$ planes respectively. The mesh is $1 \ \text{m}$ long from one end to the other. The coarse cell size is $7.5 \ \text{cm}$ and the cell sizes in the finer region vary from $0.1$ to $0.5 \ \text{cm}$. 

\begin{figure}
   \begin{subfigure}[b]{0.49\textwidth}
    \centering
    \includegraphics[width=\textwidth]{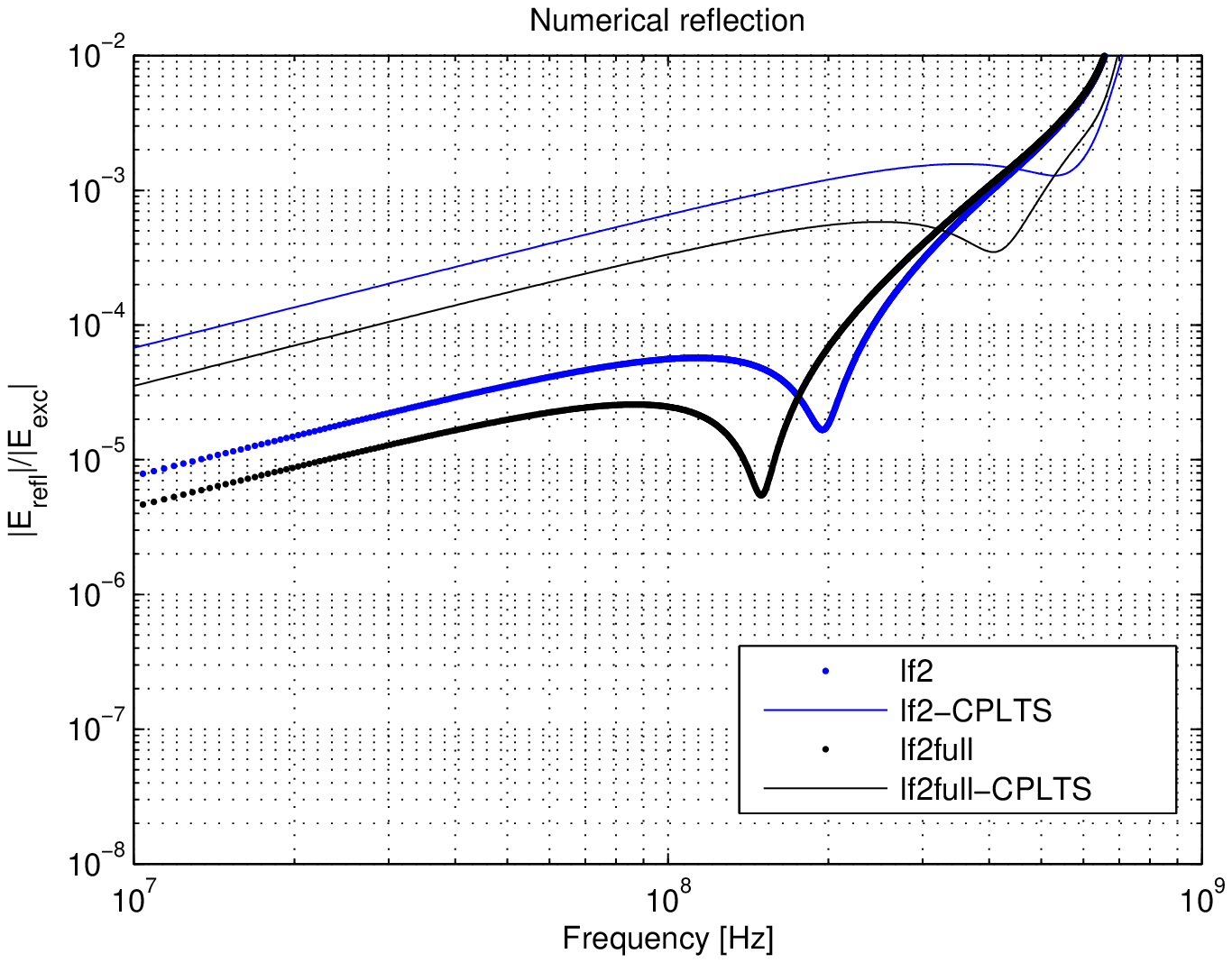}
    \caption{LF2}
  \end{subfigure}    
  \begin{subfigure}[b]{0.49\textwidth}
   \centering
   \includegraphics[width=\textwidth]{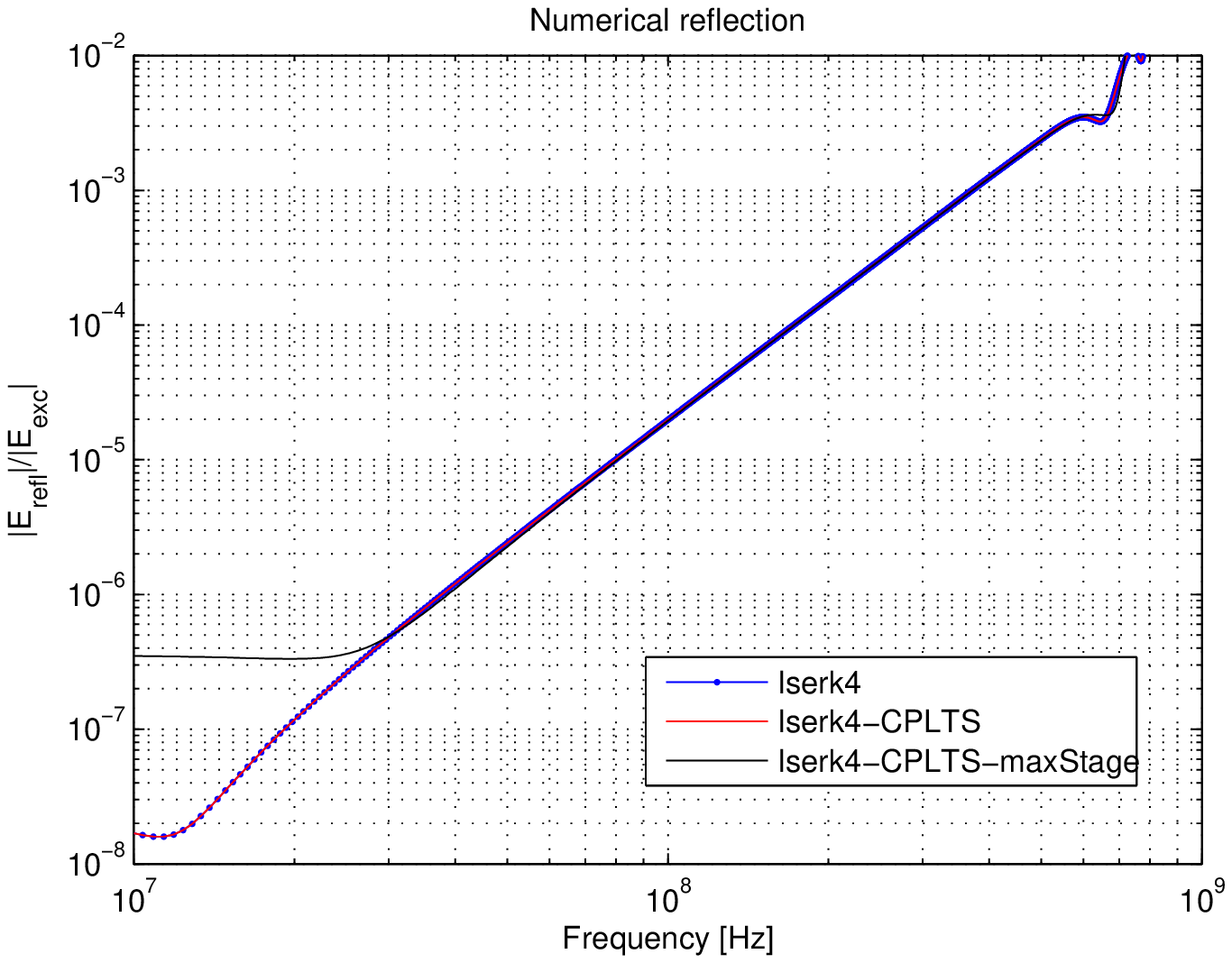}
   \caption{LSERK4}
  \end{subfigure} 
  \caption{Numerical reflection from a single interface with ratio of 15:1}
  \label{fig:refl-coeff-r15}
 
   \begin{subfigure}[b]{0.49\textwidth}
    \centering
    \includegraphics[width=\textwidth]{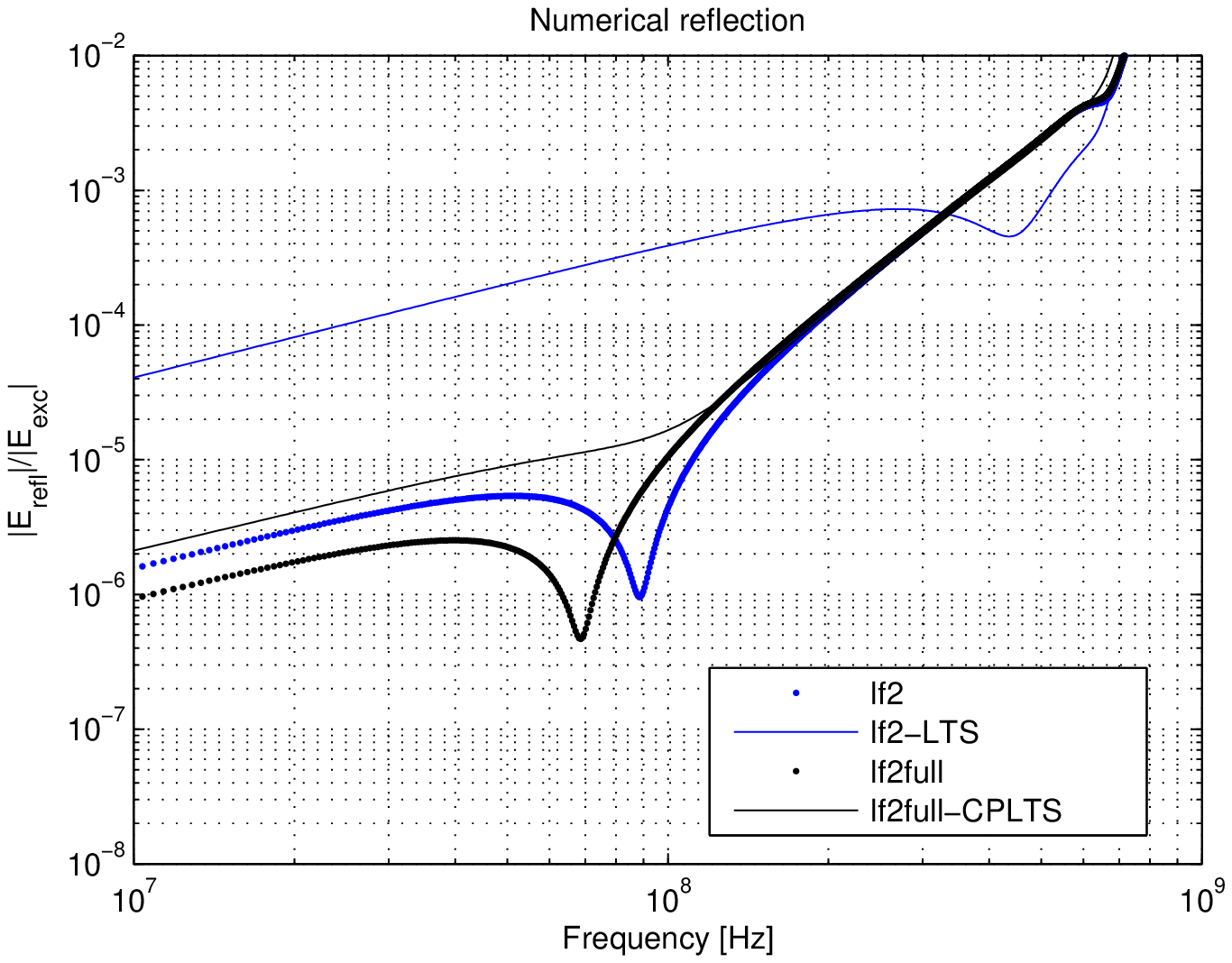}
    \caption{LF2}
  \end{subfigure}    
  \begin{subfigure}[b]{0.49\textwidth}
   \centering
   \includegraphics[width=\textwidth]{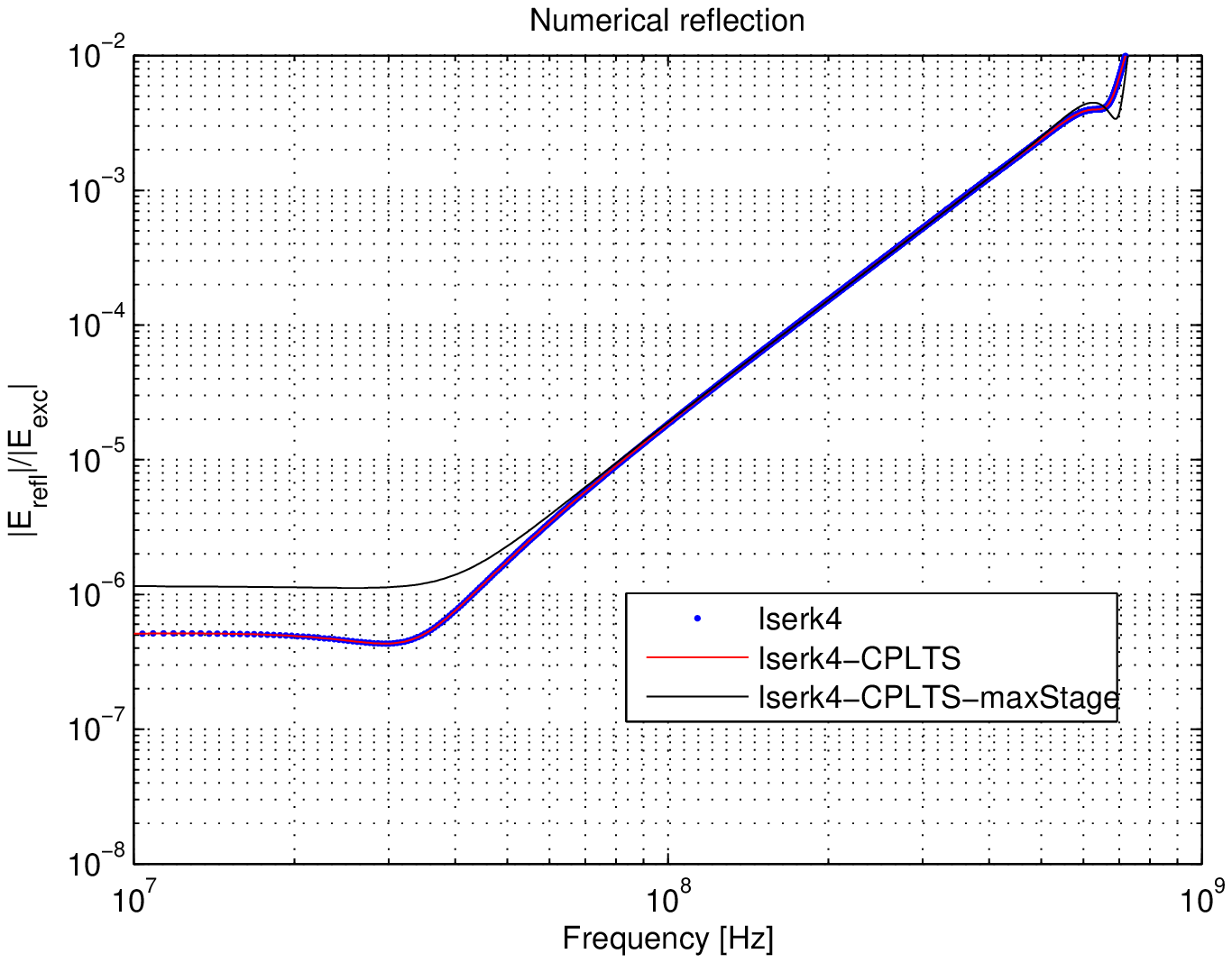}
   \caption{LSERK4}
  \end{subfigure} 
  \caption{Numerical reflection from a single interface with ratio of 75:1}
  \label{fig:refl-coeff-r75}
 
  \begin{subfigure}[b]{0.49\textwidth}
   \centering
   \includegraphics[width=\textwidth]{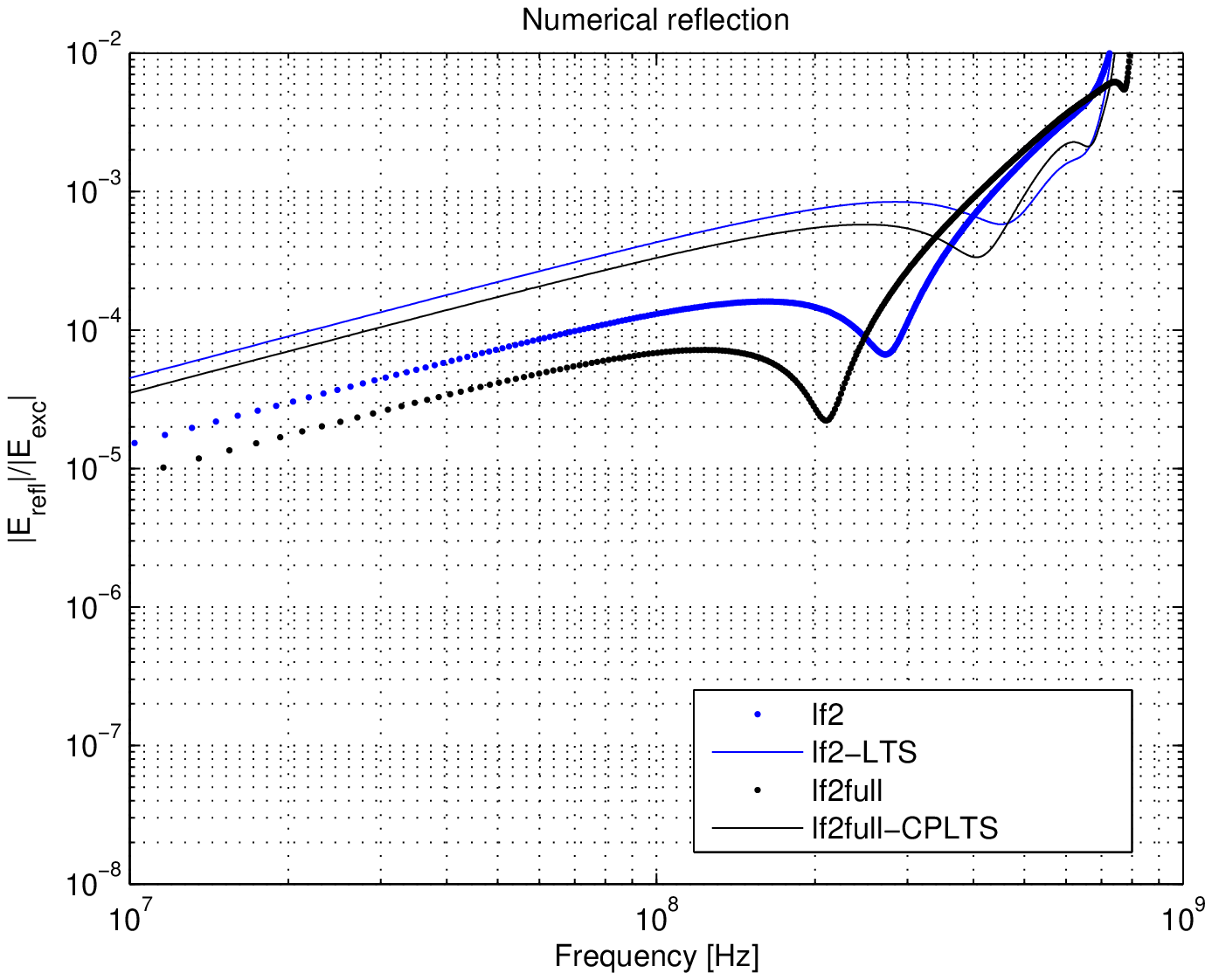}
   \caption{LF2}
  \end{subfigure}    
  \begin{subfigure}[b]{0.49\textwidth}
   \centering
   \includegraphics[width=\textwidth]{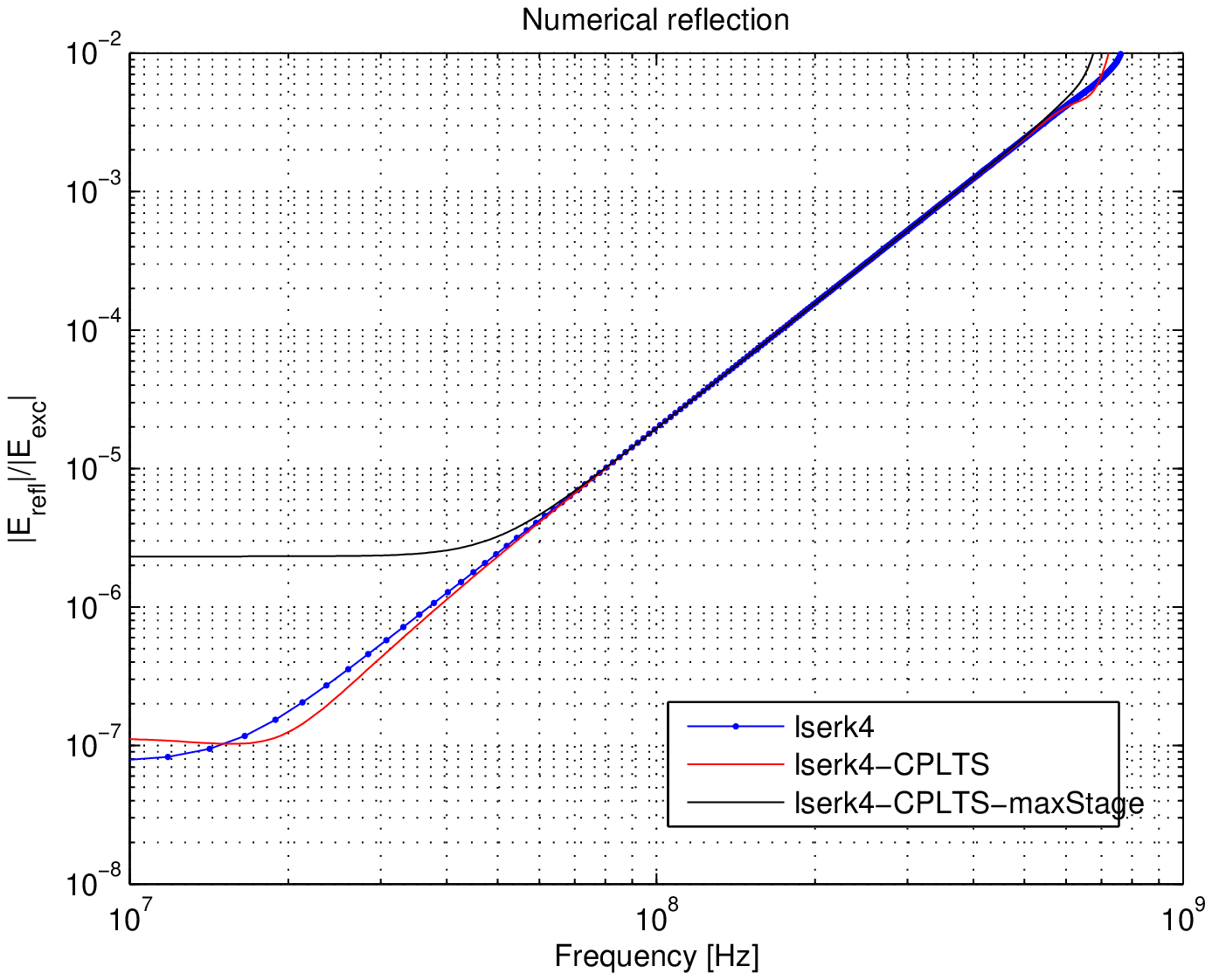}
   \caption{LSERK4}
  \end{subfigure} 
  \caption{Numerical reflection from a slab with ratio of 7.5:1.}
  \label{fig:refl-coeff-slab}
 \end{figure}

 Figures \ref{fig:refl-coeff-r15}, \ref{fig:refl-coeff-r75} and \ref{fig:refl-coeff-slab} show the reflection coefficient in a range of frequencies. The closer the values are to zero the better are the properties of the scheme.
 We observe that for this case the LF2 with a fully defined state (LF2full) exhibits slightly better properties than the classic LF2 scheme.
 A possible explanation for this is that the incident wave is resolved using more time steps.
 In LF2-LTS and LF2-CPLTS, we observe some additional degradation when compared to the classic LF2 schemes. The CPLTS exhibits less reflection than the Montseny's LTS, the difference growing with the ratio between the coarser and finer mesh. The three LSERK4 figures exhibit a better behaviour than the LF2, as expected due to the higher order of the time integration technique. 
 When the maximum stage is used for the tier assortment, we observe a higher degradation in the low-frequency regime, probably because more time--stepping operations are being performed. The results for the LSERK4-CPLTS are very encouraging as we see little differences between the use the LSERK4-CPLTS technique and the classic LSERK4.
  Table \ref{table:tierAssortingPWRefl} shows data corresponding to the tier assortment and computational times. As expected, the LF2-CPLTS is able of create more tiers than LF2-LTS because it only needs a ratio of two between maximum time step sizes. The CPU times for this simulation are listed for reference only and are not quite representative because the time employed to compute the excitation at the boundaries and the initialization is significant when compared with the operations performed to evolve the elements.
  
\begin{table}[h]
 \centering
 \begin{tabular}{ c | l | c c c c c c c | c | c }
   & Integrator & \multicolumn{7}{|c|}{Number of Elements}  &
  $\Delta t^m$ [ps]
   & CPU [s] \\
  \hline
   \multicolumn{2}{r|}{Tier} & 0 & 1 & 2 & 3 & 4 & 5 & 6      & 0 \\
    \hline
 \multirow{7}{*}{\rotatebox[origin=c]{270}{PW-Refl-r15-SInt}} 
  & LSERK4-CPLTS 
    & 120 & 312 & -  & - & - & - & - & 0.624 & 226 \\
  & LSERK4-CPLTS-mS
    & 120 & 24 & 288 & - & - & - & - & 0.624 & 485 \\
  & LSERK4
   & 432  & -  & -   & - & - & - & - & 0.624 & 468 \\
  & LF2-LTS  
   & 120 & 8 & 304 & - & - & -  & - & 0.281 &  78 \\
  & LF2     
   & 432 & -  & - & -   & - & - & - & 0.281 & 211 \\
  & LF2full-CPLTS
   & 80  & 48 & 12 & 292 & -  & - & - & 0.281 & 173 \\
  & LF2full     
   & 432 & - & - & - & -    & - & - & 0.281 & 1799 \\  
  \hline
  \multirow{7}{*}{\rotatebox[origin=c]{270}{PW-Refl-r75-SInt}}   & LSERK4-LTS 
   & 600 & 24  & 288  & -  & - & - & - & 0.12 & 3148 \\
  & LSERK4-CPLTS-mS
   & 600 & 24  & 24  & 264 & - & - & - & 0.12 & 15097 \\
  & LSERK4
   & 912 & -   & -   & -   & - & - & - & 0.12 & 4700 \\
  & LF2-LTS  
   & 600 & 8   & 12  & 292 & - & - & - & 0.06& 1444 \\
  & LF2
   & 912 & -   & -   & -   & - & - & - & 0.06 & 2296 \\
  & LF2full-CPLTS 
   & 400 & 208 & 12  & 8   & 12& 184 & 88 & 0.06 & 3524  \\
  & LF2full   
   & 912 & -   & -   & -   & - & - & - & 0.06 & 7211 \\  
  \hline
  \multirow{7}{*}{\rotatebox[origin=c]{270}{PW-Refl-r7.5-Slab}} 
  & LSERK4-CPLTS 
     & 240 & 288 & - & - & - & - & - & 1.24 & 190 \\
  & LSERK4-CPLTS-mS 
     & 240 & 288 & - & - & - & - & - & 1.24 & 342 \\     
  & LSERK4 
     & 528 & -   & - & - & - & - & - & 1.24 & 325 \\
  & LF2-LTS
    & 240 & 288  & - & - & - & - & - & 0.55  & 158 \\
  & LF2
    & 528 & -    & - & - & - & - & - & 0.55 & 151 \\
  & LF2full-CPLTS
    & 160 & 96 & 272 & - & - & - & - & 0.55 & 157 \\
  & LF2full
    & 528 & -   & - & - & - & - & -  & 0.55 & 254 \\
    \hline
 \end{tabular}
 \caption{Element Tier assorting for LTS in the plane wave reflection. \label{table:tierAssortingPWRefl}}
\end{table}
  
\subsection{PEC cavity resonances}
\begin{figure}
    \centering
   \begin{subfigure}[b]{0.9\textwidth}
    \includegraphics[width=\textwidth]{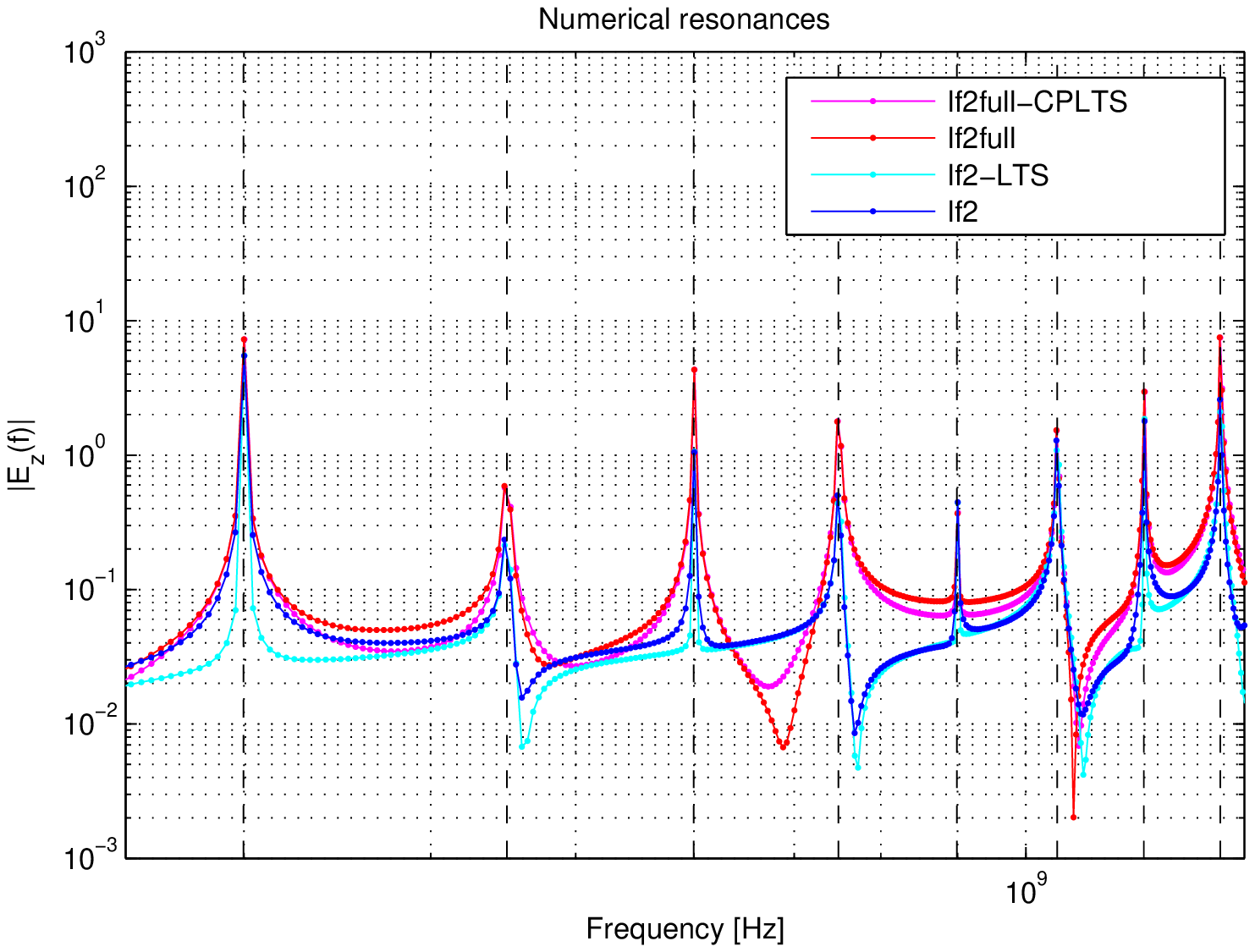}
    \caption{LF2}
   \label{fig:resonances-lf2}
  \end{subfigure}    
  \begin{subfigure}[b]{0.9\textwidth}
    \centering
    \includegraphics[width=\textwidth]{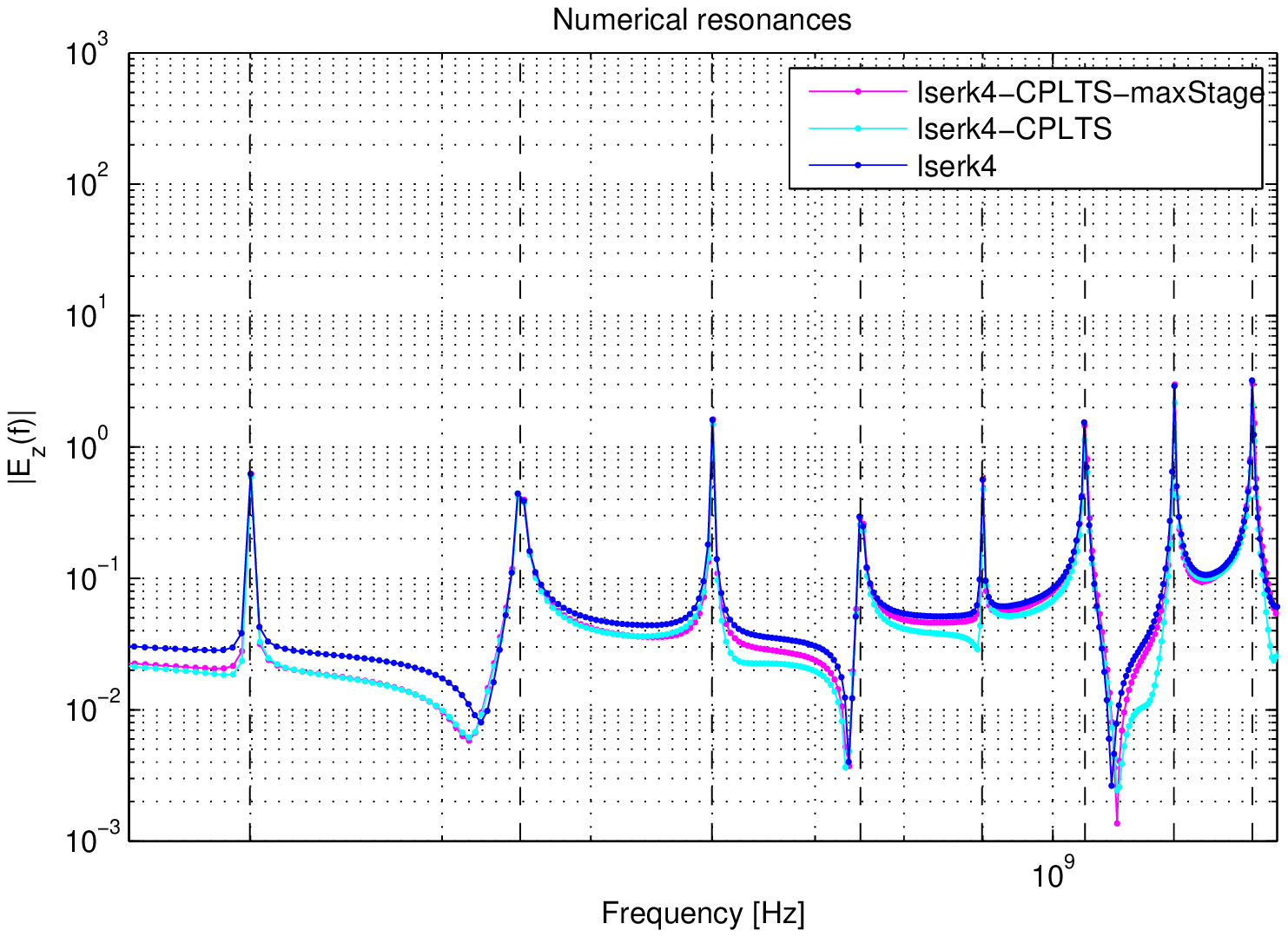}
    \caption{LSERK4}
    \label{fig:resonances-lserk4}
  \end{subfigure}    
  \caption{Resonances in a $1$ m PEC cavity with slab meshing. Vertical dashed lines represent exact eigenfrequencies.}
  \label{fig:resonances}
 \end{figure}

  As a second example we show comparisons of evolving a spatially uncorrelated random field (white noise) to study the resonances of a $1$ m PEC cavity, in a similar way as done in \cite{kim11}. The mesh used is depicted in Figure \ref{fig:slab-r7c5} with PEC boundaries at the ends rather than SMA.
The resonance frequencies are obtained by performing the Fourier transform of the electric field evolution after $250 \ \text{ns}$ at a point separated $0.3 \text{m}$ from one of the boundaries.
 Figure \ref{fig:resonances} show the eigenfrequencies obtained by the simulations together with the exact ones (black dashed vertical lines). The LF2 schemes don't show any particular difference with respect to their dispersive properties. The differences in amplitude between LF2 and LF2full can be attributed to the different initial treatment of fields.
 The LSERK4 schemes exhibit a similar behaviour in frequency but we observe additional attenuation when the CPLTS is used. When the tiers are assorted using the maximum stage criteria the attenuation is reduced.  No late time stabilities were observed in any of the simulations.
  Table \ref{table:tierAssortingResAndRCS} shows data corresponding to the tier assortment and computational times. The CPU times show a clear improvement with the LSERK4-CPLTS algorithm while the gains for the LF2-LTS are more moderate. LF2-CPLTS does not perform better than the LF2.

  \begin{table}[h]
  \centering
  \begin{tabular}{ c | l | c c c c c  | c | c }
  & Integrator & \multicolumn{5}{|c|}{Number of Elements} & $\Delta t^m$ [ps] & CPU [s] \\
  \hline
   \multicolumn{2}{r|}{Tier} & 0 & 1 & 2 & 3 & 4 & 0 \\
    \hline
 \multirow{7}{*}{\rotatebox[origin=c]{270}{reson-r7.5-Slab}} 
  & LSERK4-CPLTS 
     & 240 & 288 & - & - & -  & 1.24 & 2403 \\
  & LSERK4-CPLTS-mS 
     & 240 & 288 & - & - & -  & 1.24 & 4051 \\     
  & LSERK4 
     & 528 & -   & - & - & -  & 1.24 & 4013 \\
  & LF2-LTS
    & 240 & 288  & - & - & -  & 0.55 & 1207 \\
  & LF2
    & 528 & -    & - & - & -  & 0.55 & 1917 \\
  & LF2full-CPLTS
    & 160 & 96 & 272 & - & -  & 0.55 & 2381  \\
  & LF2full
    & 528 & -   & - & - & -   & 0.55 & 3615 \\
   \hline
    \multirow{6}{*}{\rotatebox[origin=c]{270}{rcs-1m$\*$}} 
  & LSERK4-CPLTS 
     & 4535 & 57279 & 157 & - & -  & 2.1 & 3733 \\  
  & LSERK4 
     & 61971 & -   & - & - & - & 2.1 & 8613 \\
  & LF2-LTS
    & 522 & 8614 & 52798 & 37  & - & 0.95 & 963 \\
  & LF2
    & 61971 & -    & - & -  & - & 0.95 & 4348 \\
  & LF2full-CPLTS
    & 114 & 2155 & 7411 & 34521 & 17770  &  0.95 & 1851  \\
  & LF2full
    & 61971 & -   & - & - & -  & 0.95 &  8642 \\
   \hline
 \end{tabular}
 \caption{Element Tier assorting for LTS in the resonant cavity and RCS problems.     
 \label{table:tierAssortingResAndRCS}}
\end{table}

 \subsection{RCS Analysis of a PEC Sphere}
\begin{figure}
  \centering
  \begin{subfigure}[b]{0.3\textheight}
   \centering
   \includegraphics[width=\textwidth]{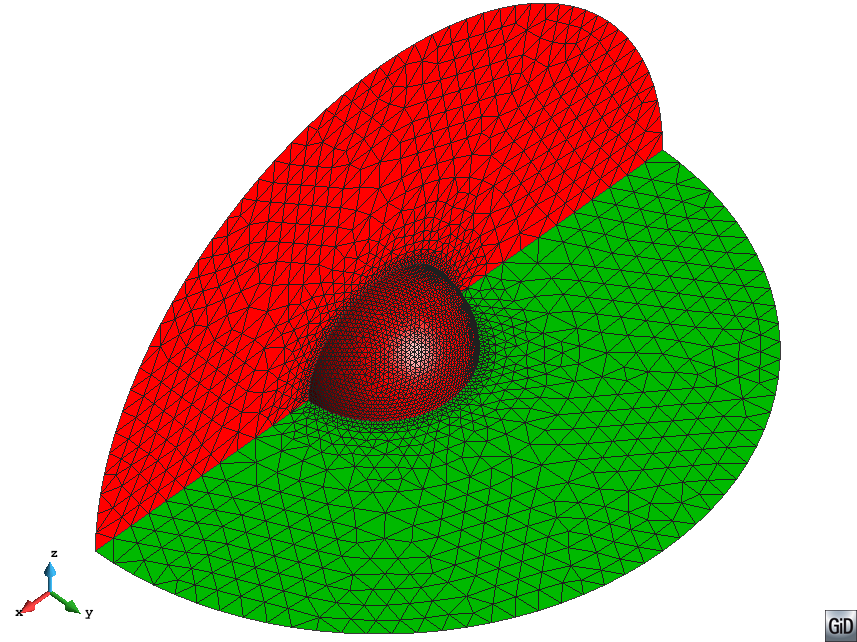}
  \caption{PEC (red) and PMC (green). SMA is not depicted.}
  \label{fig:rcs-pec-pmc}
  \end{subfigure}    
  \begin{subfigure}[b]{0.3\textheight}
   \centering
   \includegraphics[width=\textwidth]{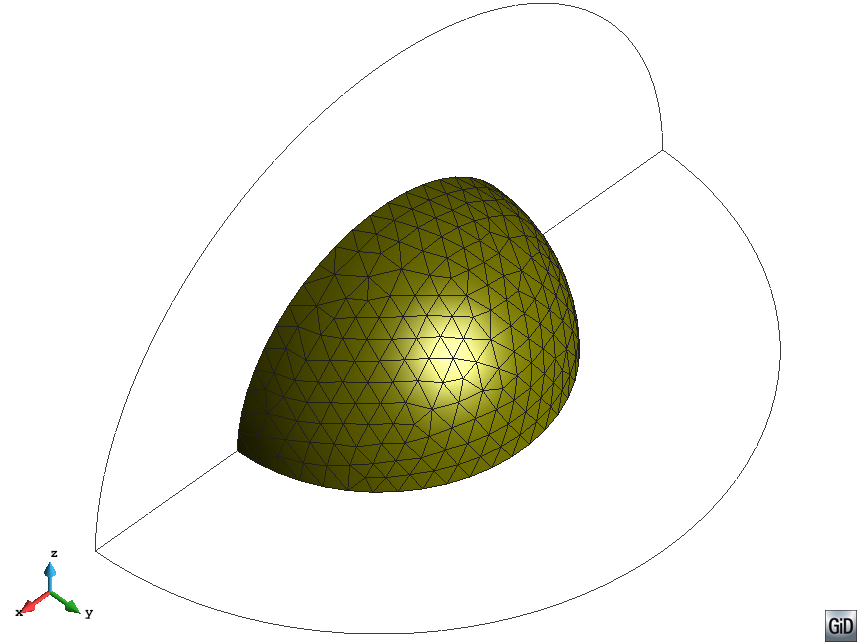}
   \caption{Total Field region}
   \label{fig:rcs-tfsf}
  \end{subfigure} 
  \caption{Boundary conditions for the RCS case.}
  \label{fig:rcs-bc}
 \end{figure}

 As a last test case we present a bi--static Radar Cross Section (RCS) analysis \cite{alvarez10}.
 Figure \ref{fig:rcs-bc} show the boundary conditions used. Symmetry conditions were used to reduce the computational domain and the $1 \ \text{m}$ radius sphere was modelled using a PEC boundary condition. SMA boundary conditions were used to terminate the domain $3  \ \text{m}$ away from the surface of the sphere. The illumination was done using a Total Field/Scattered Field boundary condition in a spherical surface located $1  \ \text{m}$ away from the sphere using a Gaussian wave with $1  \ \text{ns}$ spread, $y$-polarization and propagating along the $x$ axis. The typical element size of the mesh was $25  \ \text{cm}$ everywhere except in the PEC spherical surface modelling the sphere in which was set to $5 \ \text{cm}$.

\begin{figure}
  \begin{subfigure}[b]{0.49\textwidth}
   \centering
   \includegraphics[width=\textwidth]{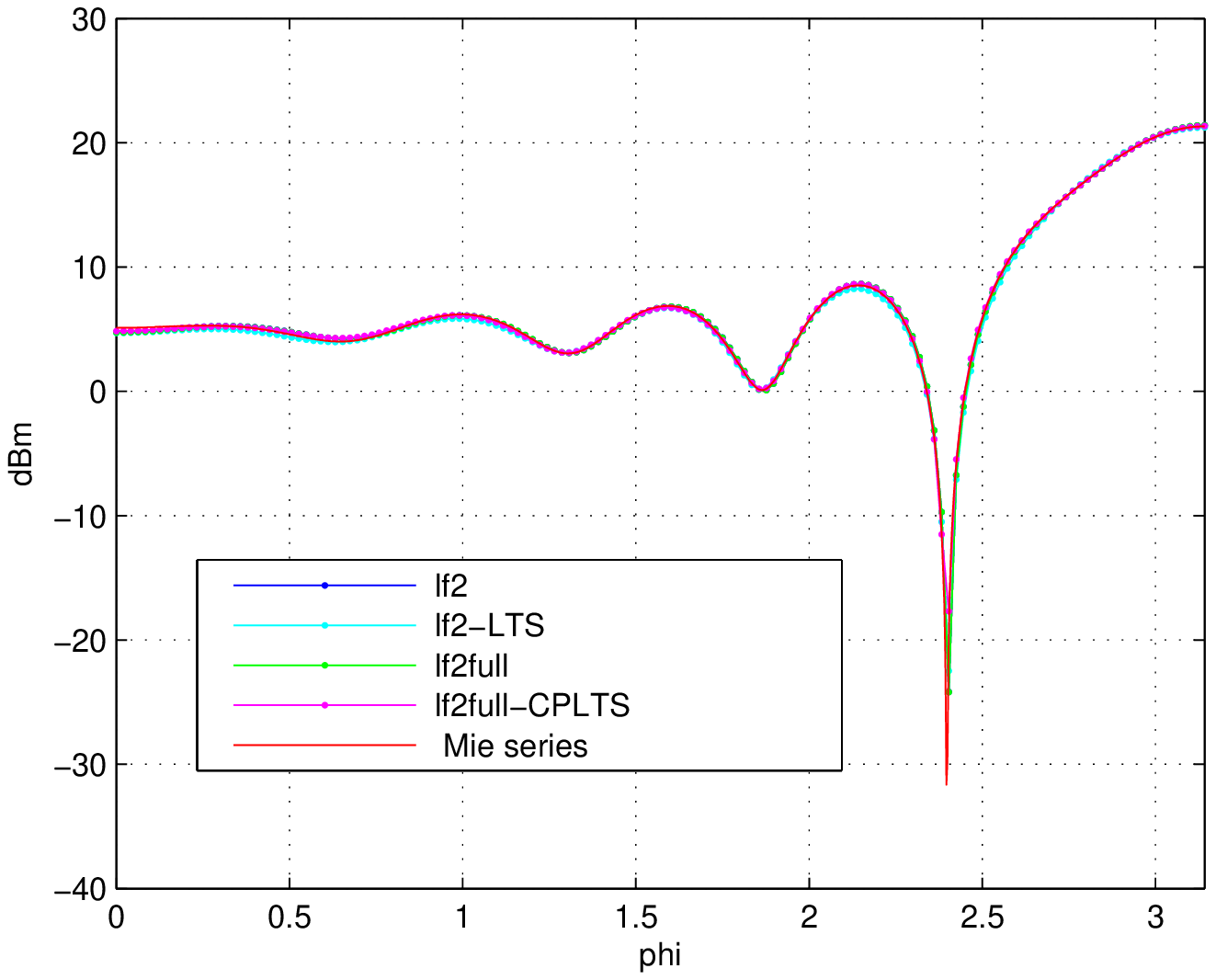}
   \caption{LF2}
   \label{fig:rcs-lf2-300MHz}
  \end{subfigure}    
  \begin{subfigure}[b]{0.49\textwidth}
   \centering
   \includegraphics[width=\textwidth]{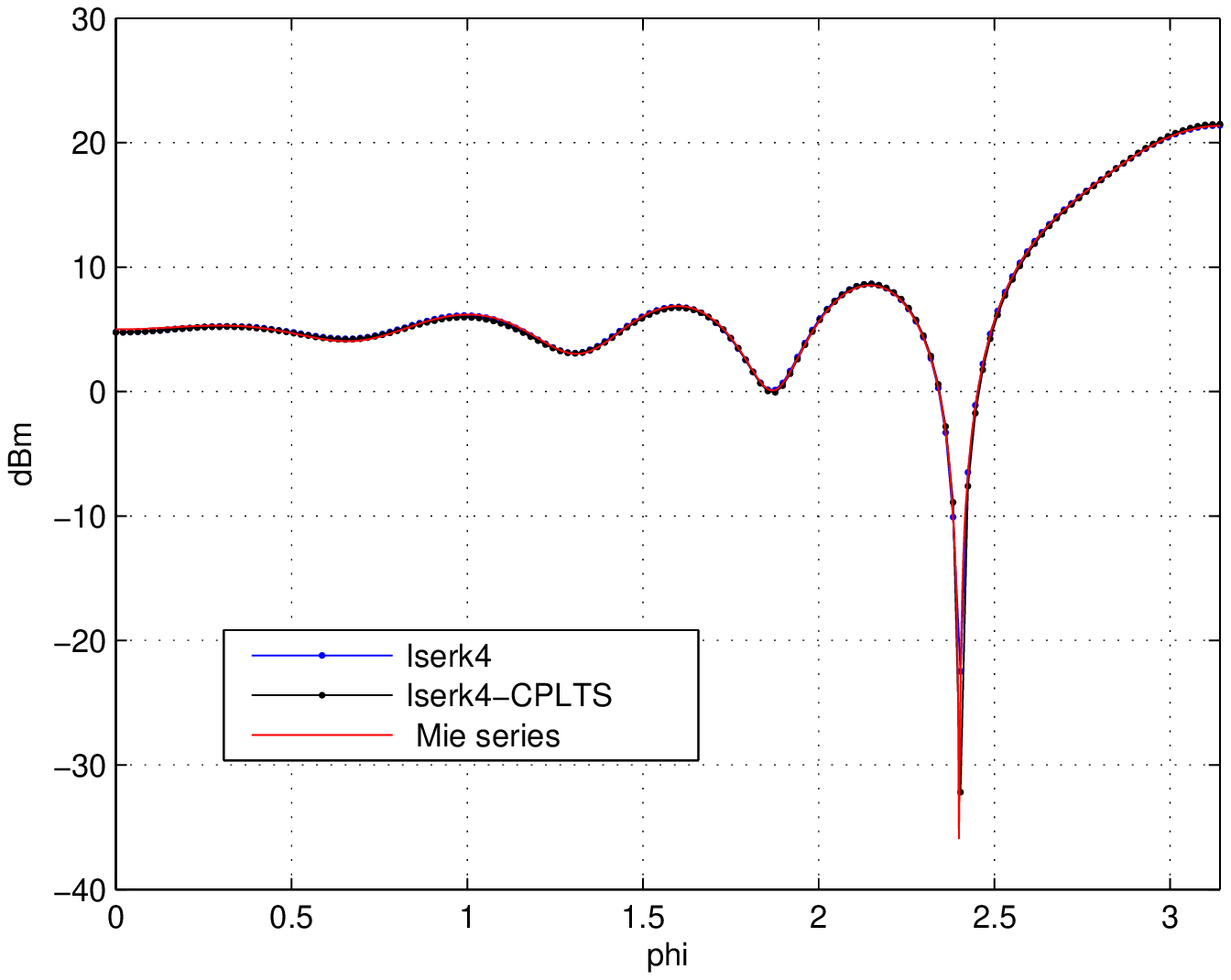}
   \caption{LSERK4}
   \label{fig:rcs-rk-300MHz}
  \end{subfigure} 
  
  \begin{subfigure}[b]{0.49\textwidth}
   \centering
   \includegraphics[width=\textwidth]{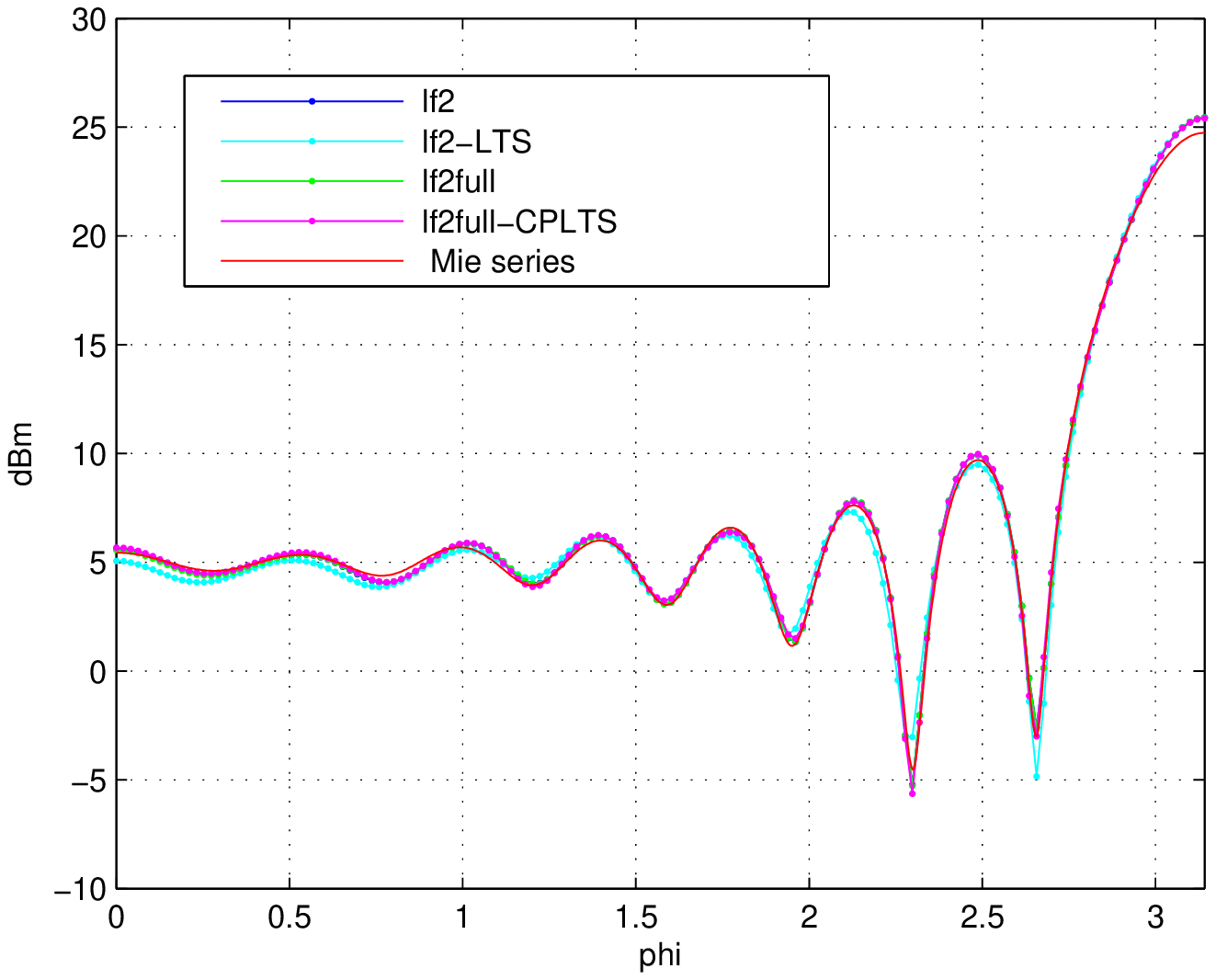}
   \caption{LF2}
   \label{fig:rcs-lf2-450MHz}
  \end{subfigure}    
  \begin{subfigure}[b]{0.49\textwidth}
   \centering
   \includegraphics[width=\textwidth]{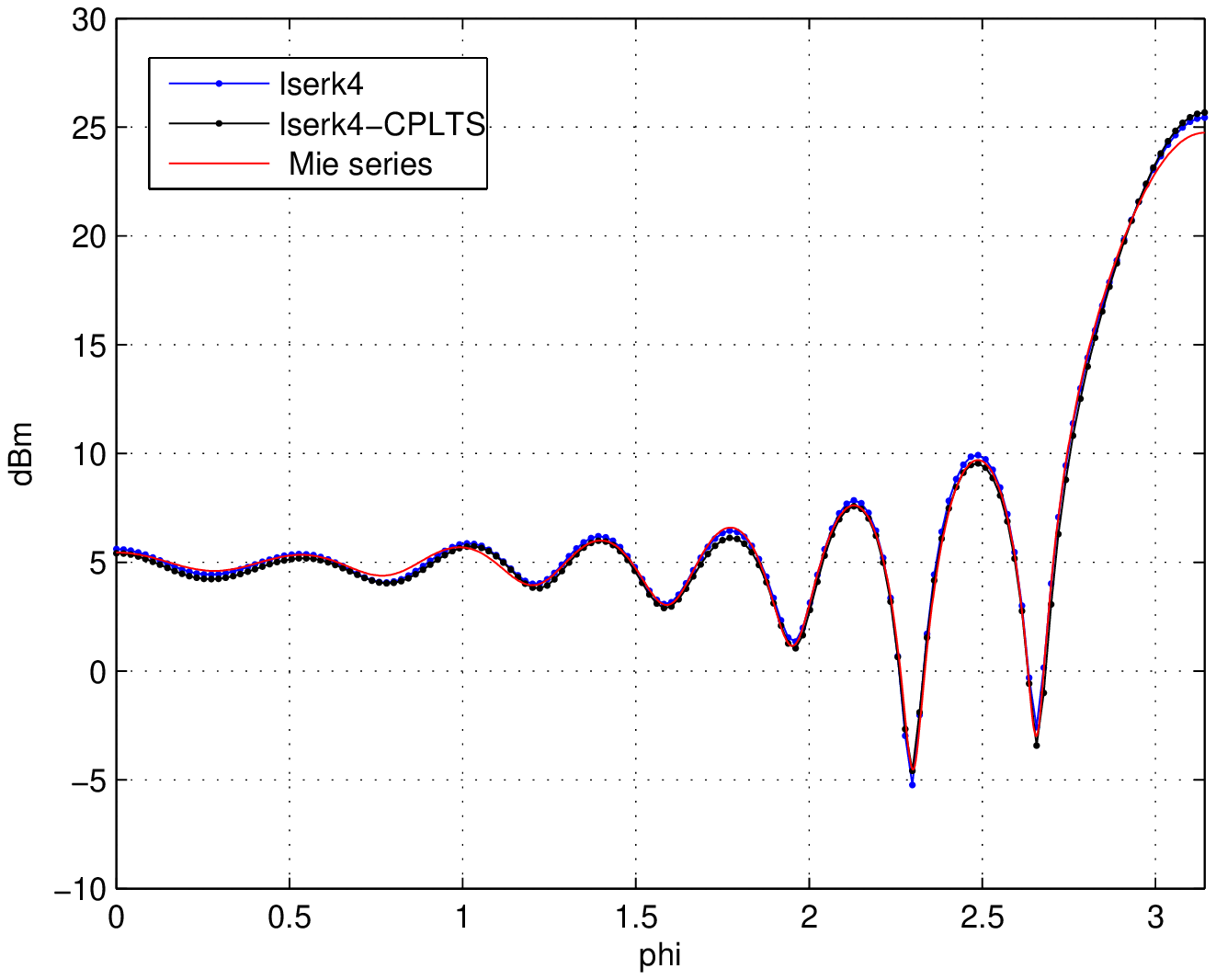}
   \caption{LSERK4}
   \label{fig:rcs-rk-450MHz}
  \end{subfigure} 

  \begin{subfigure}[b]{0.49\textwidth}
   \centering
   \includegraphics[width=\textwidth]{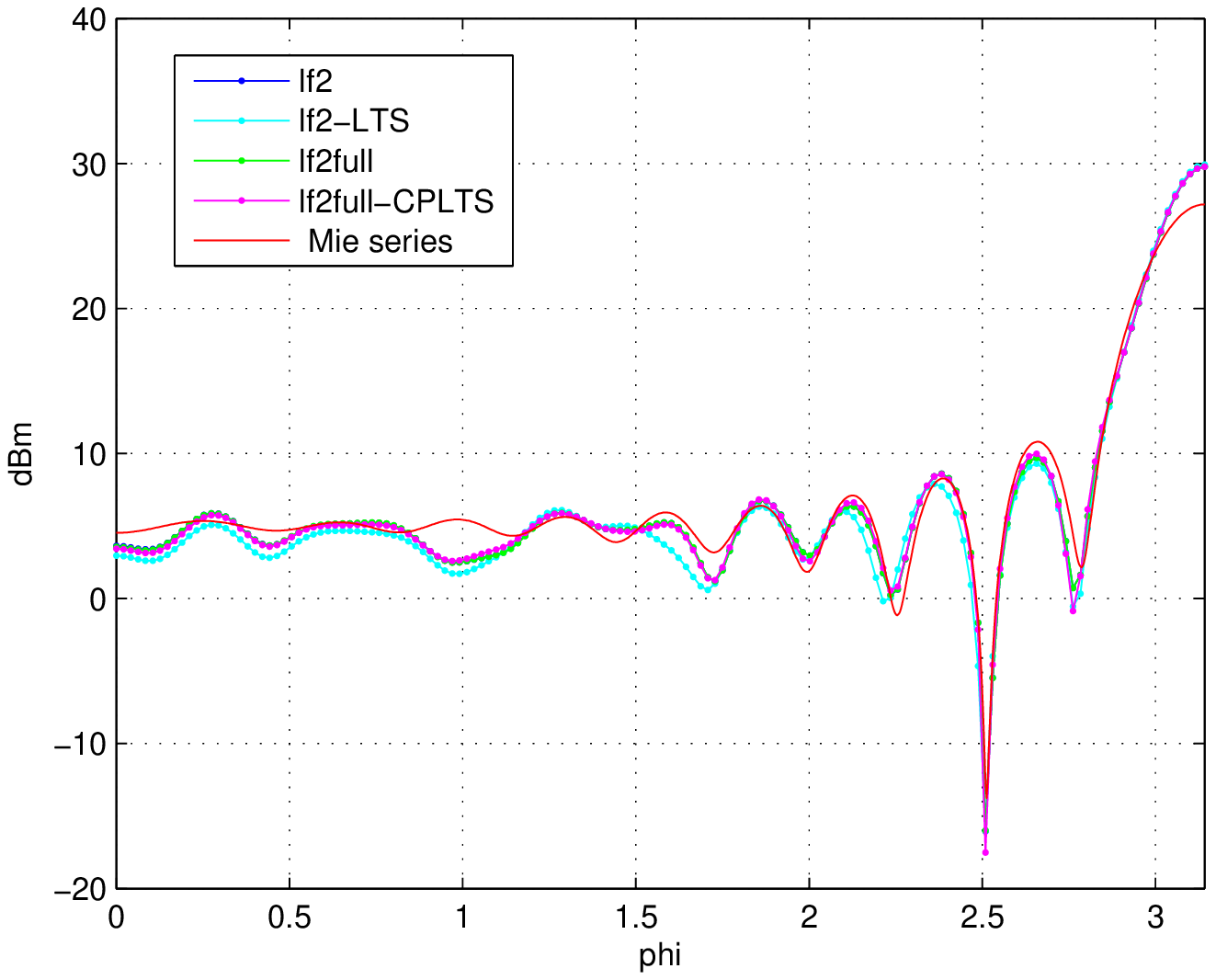}
   \caption{LF2}
   \label{fig:rcs-lf2-600MHz}
  \end{subfigure}    
  \begin{subfigure}[b]{0.49\textwidth}
   \centering
   \includegraphics[width=\textwidth]{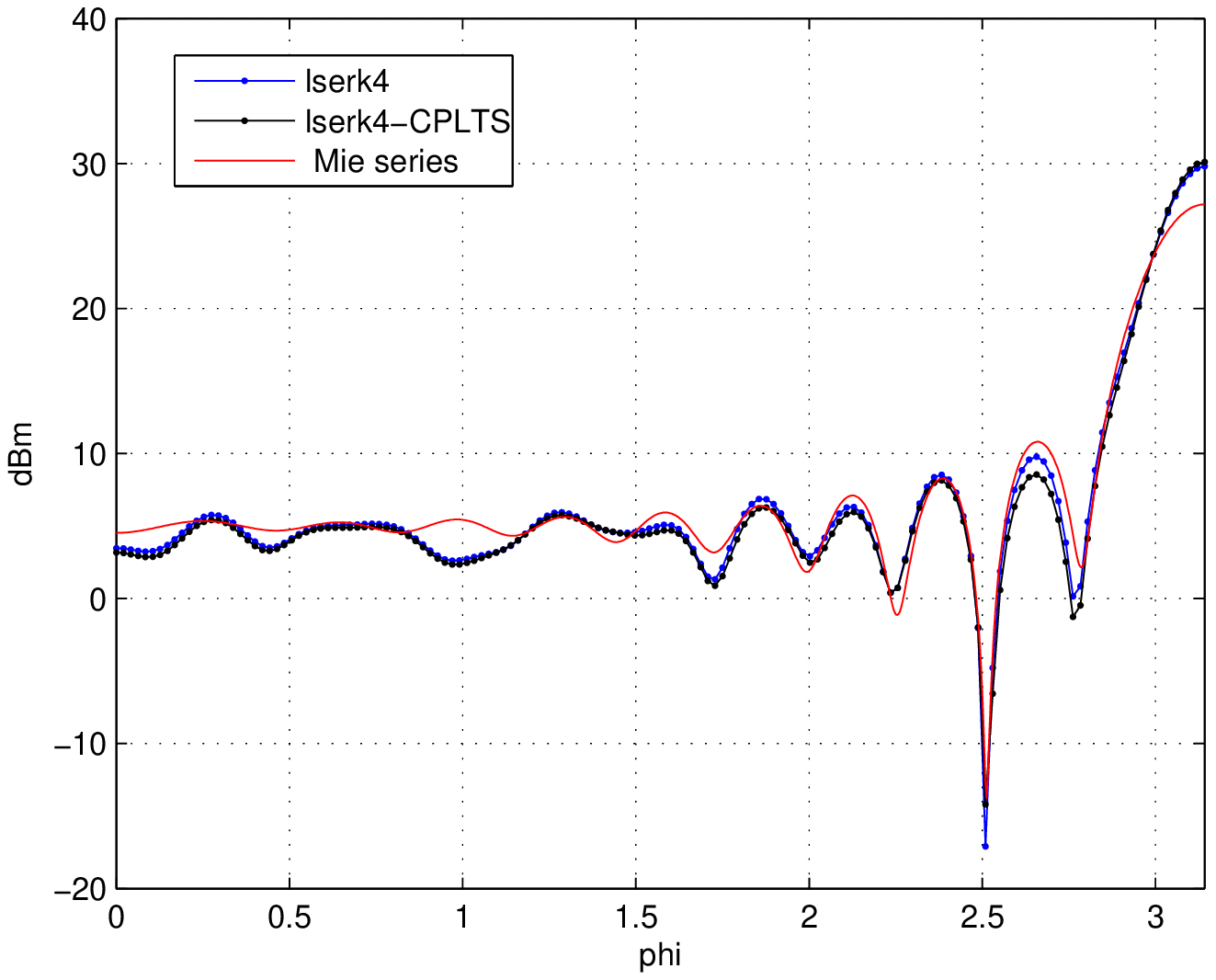}
   \caption{LSERK4}
   \label{fig:rcs-rk-600MHz}
  \end{subfigure} 
  \caption{Bi-static RCS at $300 \ \text{MHz}$ (top row), $450 \ \text{MHz}$ (middle row) and $600 \ \text{MHz}$ (bottom row). Continuous red line represents the analytical solution obtained through Mie's series.}
  \label{fig:rcs}

 \end{figure}

 Figure \ref{fig:rcs} shows the results of the analysis for the various LF2 and LSERK4 schemes under study. At $450 \ \text{MHz}$ we see that the LF2 methods fit the Mie's analytical solution but the LF2 using Montseny's approach exhibits an angular offset caused by an appreciable difference in the dispersion relation. At $600 \text{MHz}$ all methods present a higher deviation, caused by a poorer resolution of the spatial grid.
 
 The LSERK4 results exhibit a better behaviour than the LF2, capturing the main features of the analytical solution. The application of CPLTS seems to better preserve the dispersion relation an thus the position of the peaks. However, at $600 \ \text{MHz}$ we can observe an appreciable numerical dissipation being introduced.
 
 Table \ref{table:tierAssortingResAndRCS} shows data corresponding to the tier assortment and computational times. In this case, the LSERK4-CPLTS is able to provide a considerable speed up, reducing the CPU time from $8613$ to $3733 \ \text{s}$ ($\sim 2$). The LF2 LTS techniques yield a speed--up of about four times the non-LTS counterparts. The CPLTS speeds up the classic LF2 by a factor about two.
 
 \section{Tier assortment}
\begin{figure}
  \begin{subfigure}[b]{0.49\textwidth}
   \centering
   \includegraphics[width=\textwidth]{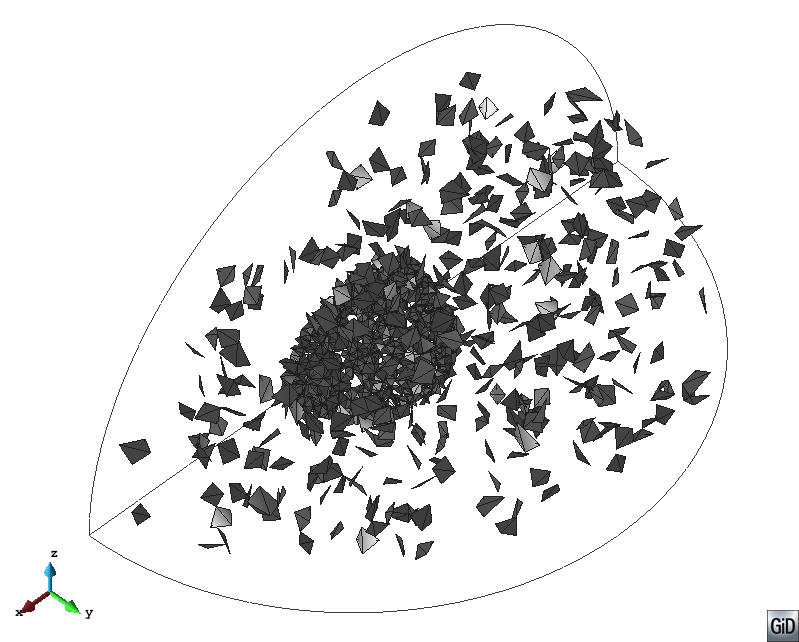}
   \caption{Tier 0}
  \end{subfigure}    
  \begin{subfigure}[b]{0.49\textwidth}
   \centering
   \includegraphics[width=\textwidth]{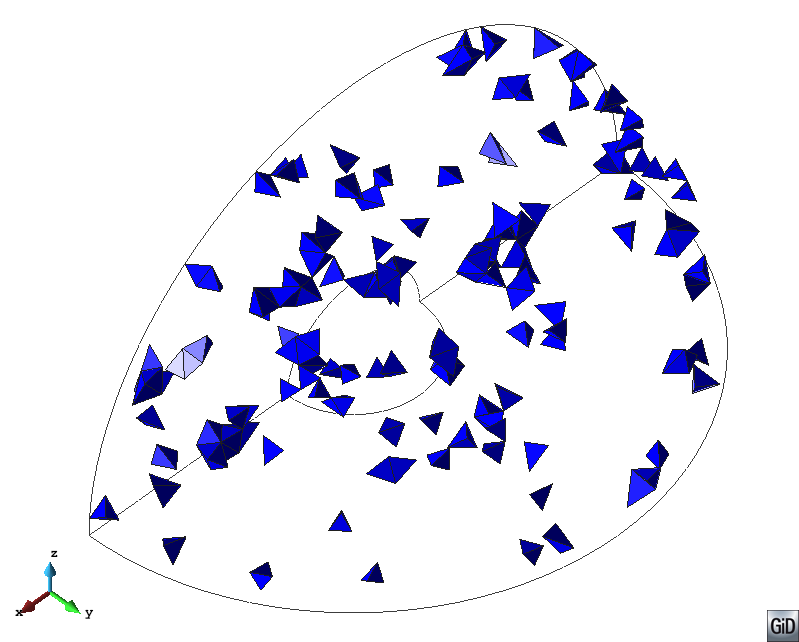}
   \caption{Tier 2}
  \end{subfigure} 
  \caption{Tier assortment for LSERK4. Tier 1 is not represented.}
  \label{fig:rcs-rk-tierassortment}
  
  \begin{subfigure}[b]{0.49\textwidth}
   \centering
   \includegraphics[width=\textwidth]{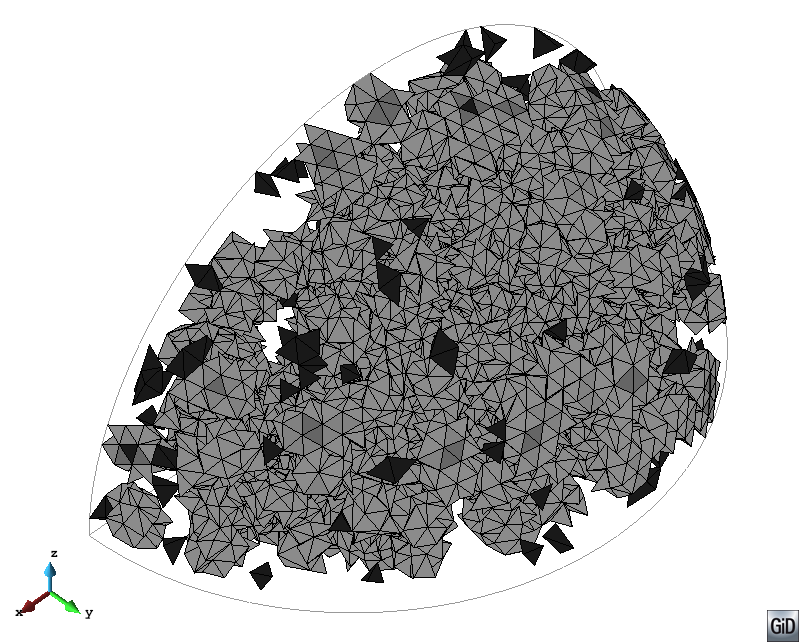}
   \caption{Stages distribution. Isometric.}
  \end{subfigure}    
  \begin{subfigure}[b]{0.49\textwidth}
   \centering
   \includegraphics[width=\textwidth]{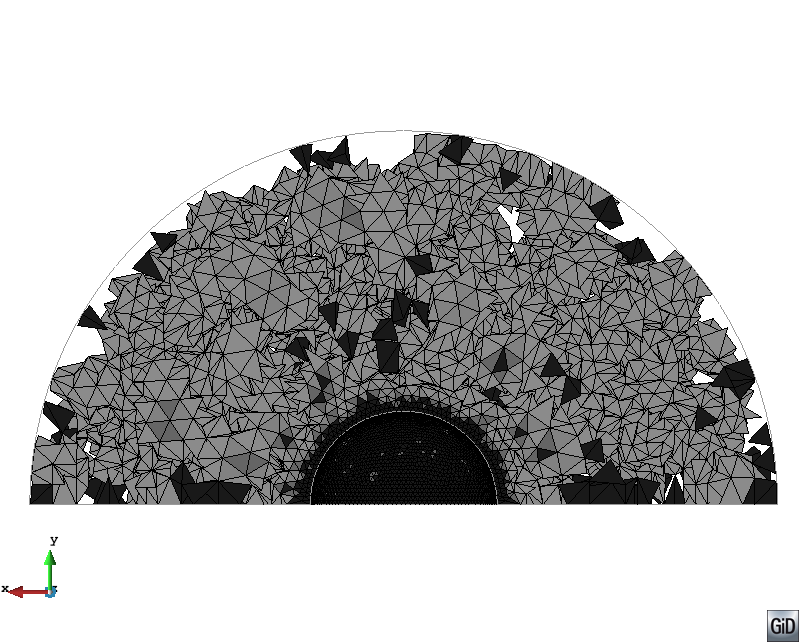}
   \caption{Stages distribution. XY plane.}
  \end{subfigure} 
  \caption{Elements in LSERK4 where some operations are required by the smaller tiers. Darker colour means more operations (closer to a smaller tier). The degrees of freedom belonging to the elements represented need to be stored when the smaller tier is solved. Elements that do not require additional operations and storage are not represented.}
  \label{fig:rcs-rk-stageassortment}
 \end{figure}

\begin{figure}
  \begin{subfigure}[b]{0.49\textwidth}
   \centering
   \includegraphics[width=\textwidth]{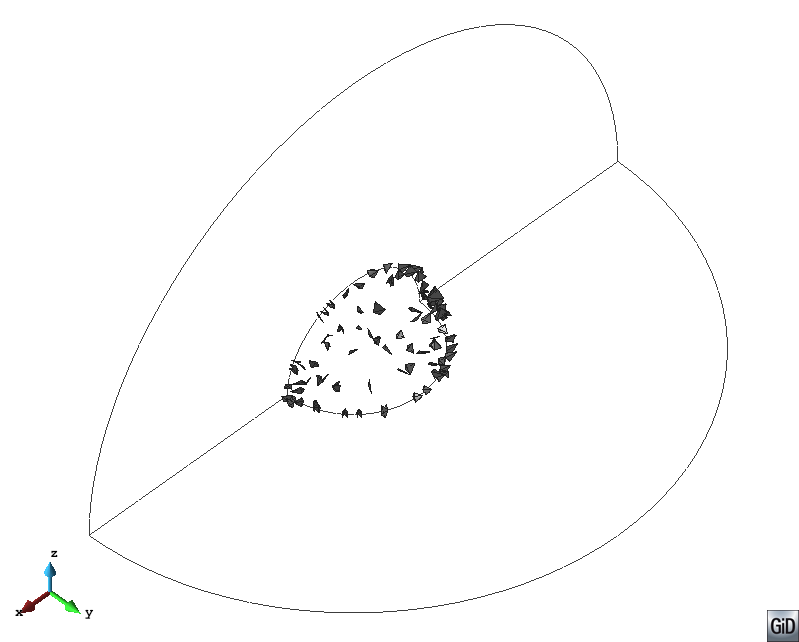}
   \caption{Tier 0}
  \end{subfigure}    
  \begin{subfigure}[b]{0.49\textwidth}
   \centering
   \includegraphics[width=\textwidth]{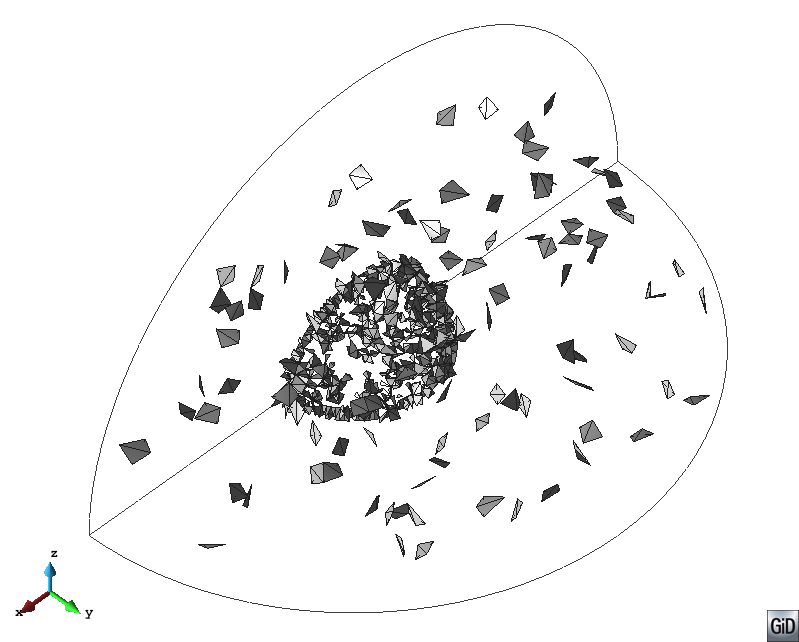}
   \caption{Tier 1}
  \end{subfigure} 
  
  \begin{subfigure}[b]{0.49\textwidth}
   \centering
   \includegraphics[width=\textwidth]{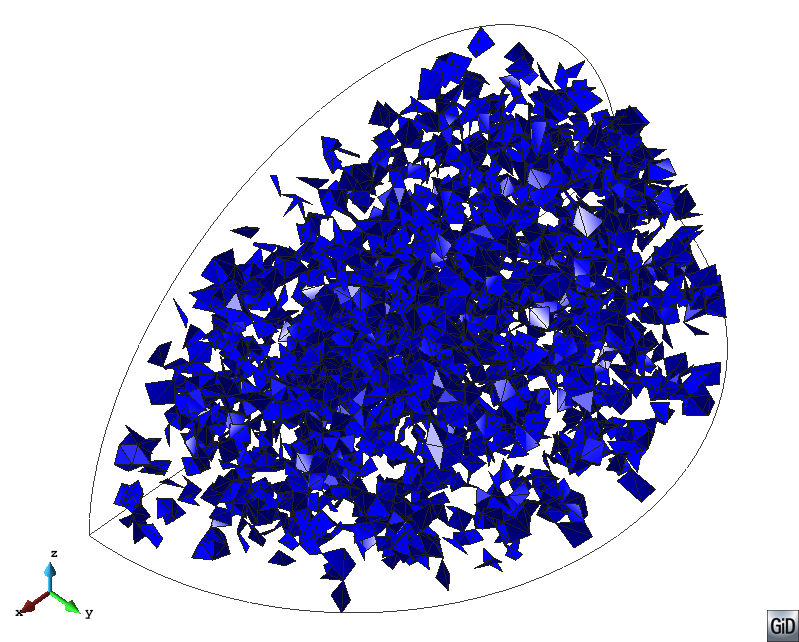}
   \caption{Tier 2}
  \end{subfigure}    
  \begin{subfigure}[b]{0.49\textwidth}
   \centering
   \includegraphics[width=\textwidth]{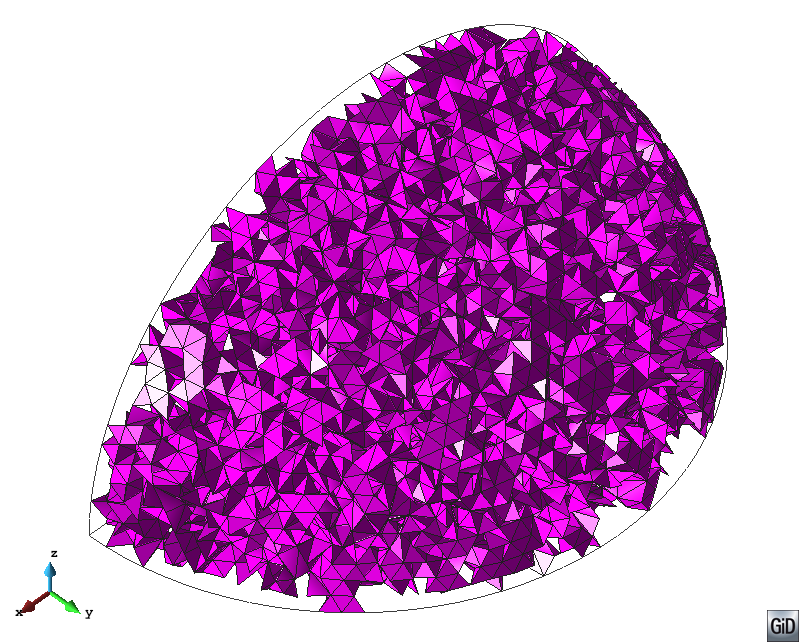}
   \caption{Tier 3}
  \end{subfigure} 
  
  \centering
  \begin{subfigure}[b]{0.49\textwidth}
   \centering
   \includegraphics[width=\textwidth]{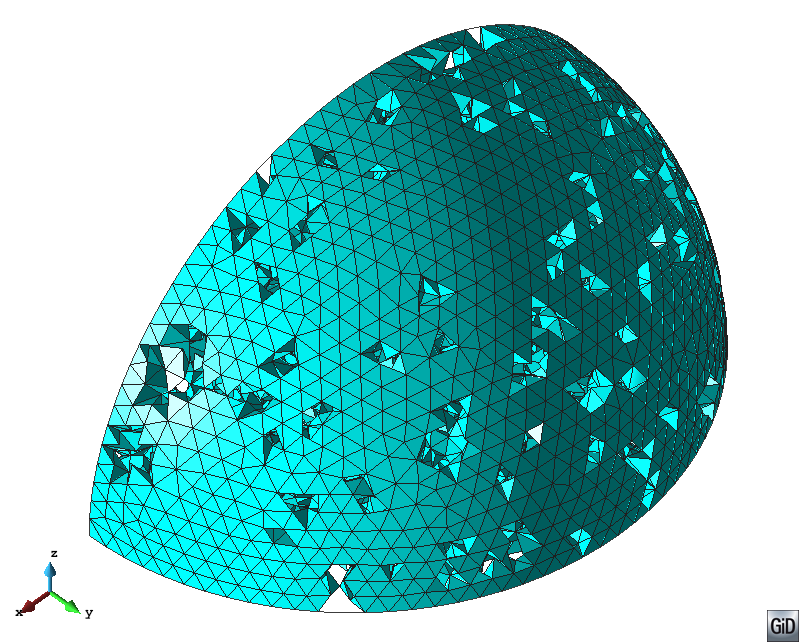}
   \caption{Tier 4}
  \end{subfigure} 
  \caption{Tier assortment for LF2-CPLTS.}
  \label{fig:rcs-lf2full-tierassortment}
 \end{figure}

\begin{figure}
  \begin{subfigure}[b]{0.49\textwidth}
   \centering
   \includegraphics[width=\textwidth]{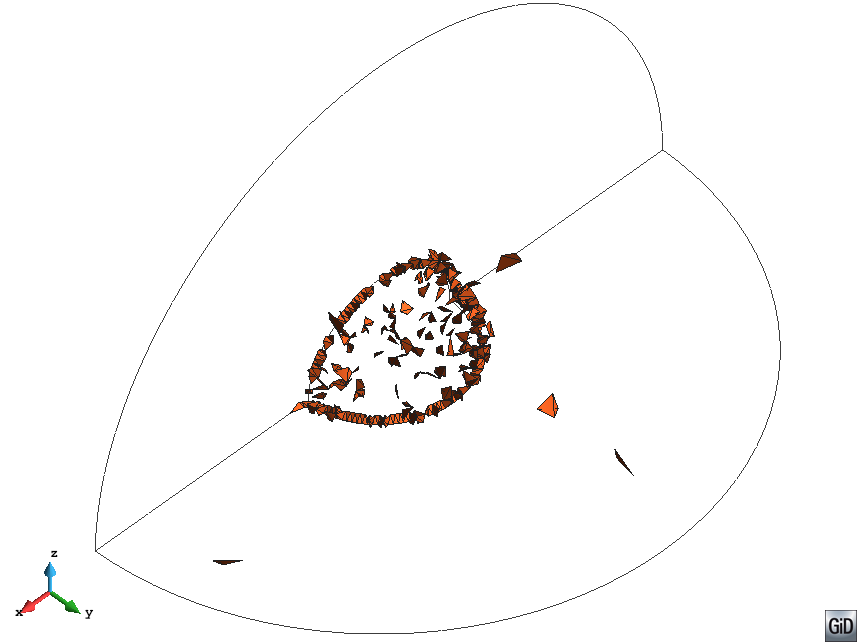}
   \caption{Tier 0}
  \end{subfigure}    
  \begin{subfigure}[b]{0.49\textwidth}
   \centering
   \includegraphics[width=\textwidth]{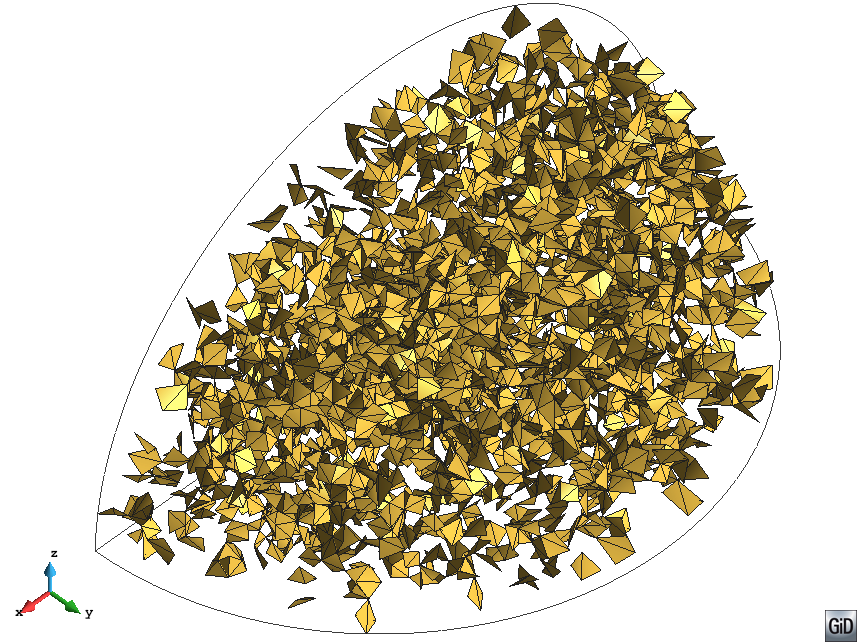}
   \caption{Tier 1}
  \end{subfigure} 
  
  \begin{subfigure}[b]{0.49\textwidth}
   \centering
   \includegraphics[width=\textwidth]{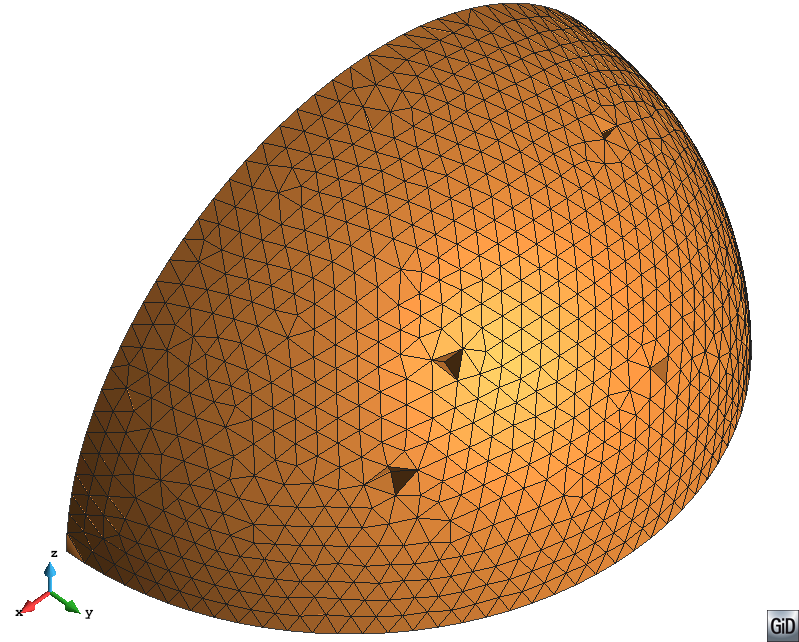}
   \caption{Tier 2}
  \end{subfigure}    
  \begin{subfigure}[b]{0.49\textwidth}
   \centering
   \includegraphics[width=\textwidth]{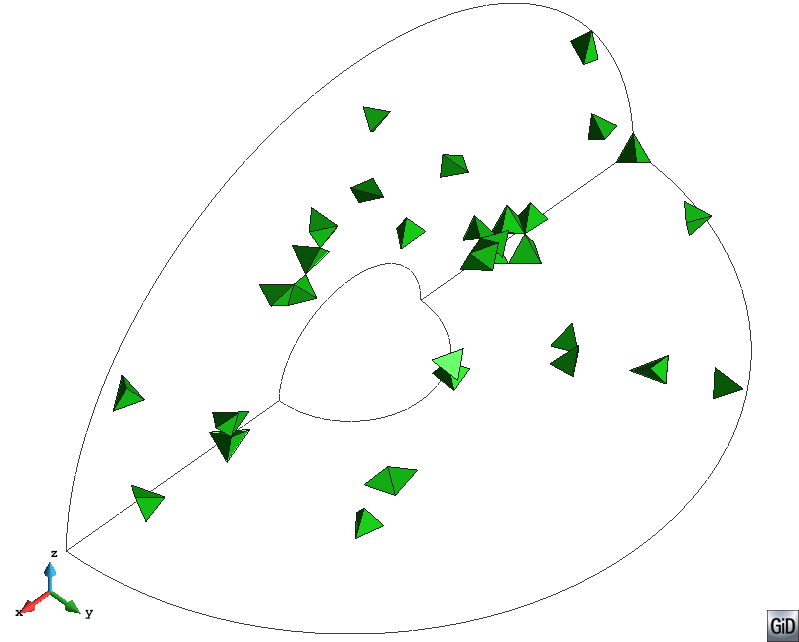}
   \caption{Tier 3}
  \end{subfigure} 
  \caption{Tier assortment for LF2-LTS.}
  \label{fig:rcs-lf2-tierassortment}
 \end{figure}

\begin{figure}
  \begin{subfigure}[b]{0.49\textwidth}
   \centering
   \includegraphics[width=\textwidth]{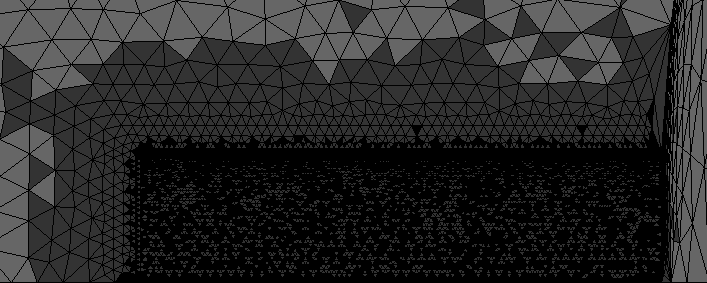}
   \caption{LSERK4-CPLTS Detail view}
  \end{subfigure}    
  \begin{subfigure}[b]{0.49\textwidth}
   \centering
   \includegraphics[width=\textwidth]{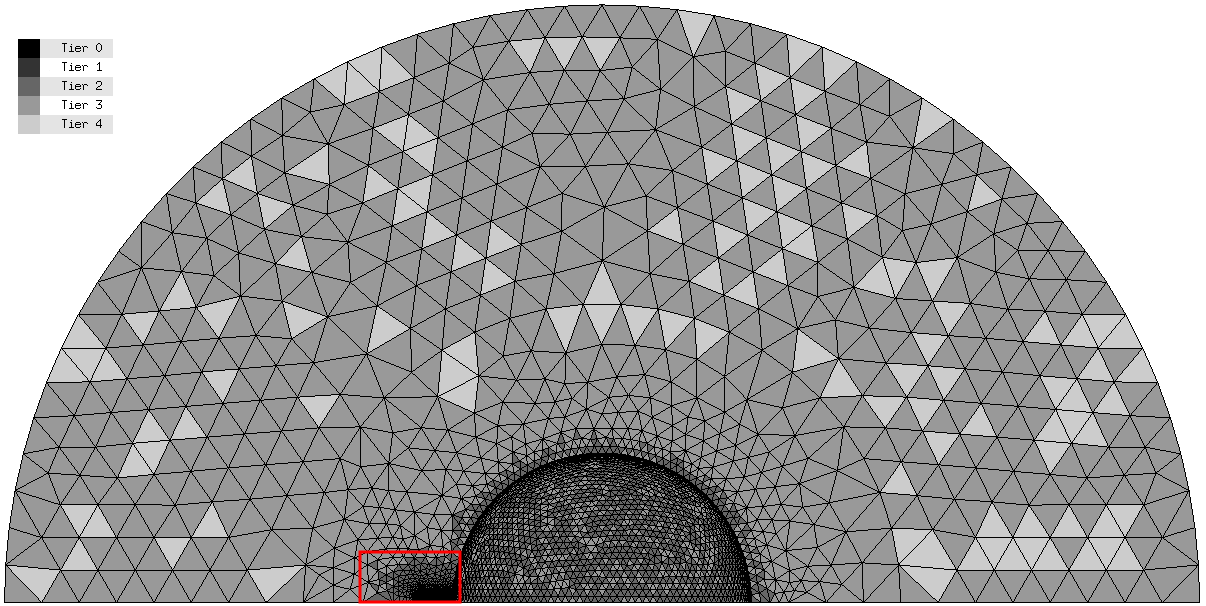}
   \caption{LSERK4-CPLTS General view}
  \end{subfigure}    
  \caption{Tier assortment for LSERK4-CPLTS}
  \label{fig:antenna-rk-tierassortment}

  \begin{subfigure}[b]{0.49\textwidth}
   \centering
   \includegraphics[width=\textwidth]{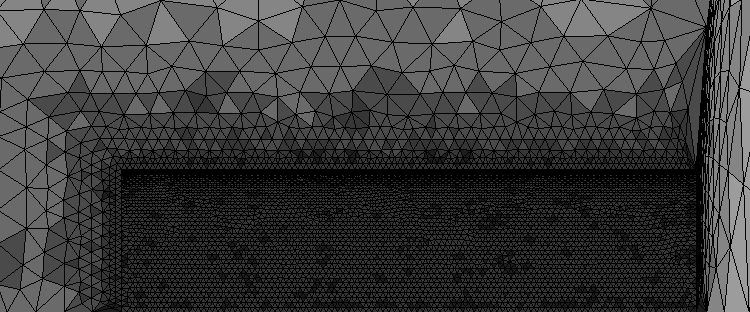}
   \caption{LF2-CPLTS Detail view}
  \end{subfigure}    
  \begin{subfigure}[b]{0.49\textwidth}
   \centering
   \includegraphics[width=\textwidth]{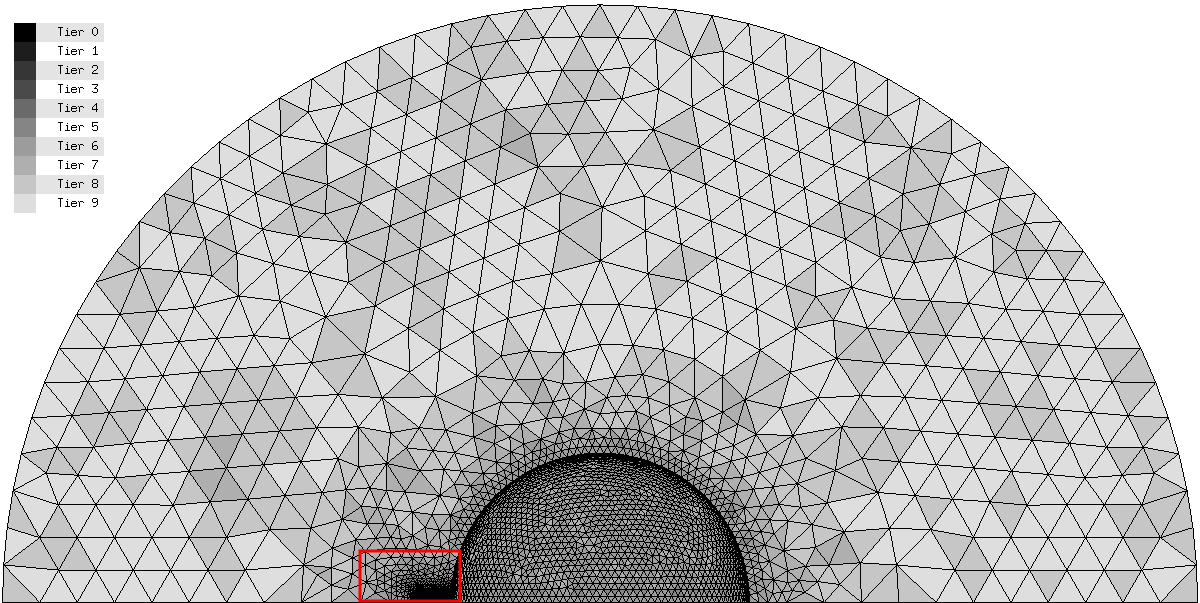}
   \caption{LF2-CPLTS General view}
  \end{subfigure}    
  \caption{Tier assortment for LF2-CPLTS}
  \label{fig:antenna-lf2full-tierassortment}
  
  \begin{subfigure}[b]{0.49\textwidth}
   \centering
   \includegraphics[width=\textwidth]{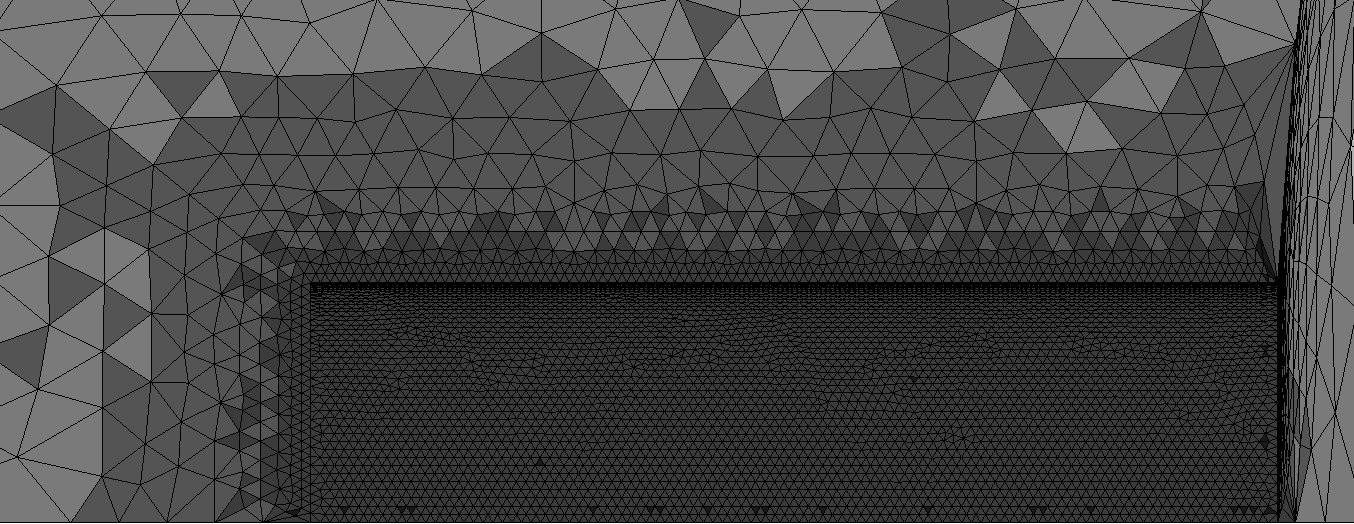}
   \caption{LF2-LTS Detail view}
  \end{subfigure}    
  \begin{subfigure}[b]{0.49\textwidth}
   \centering
   \includegraphics[width=\textwidth]{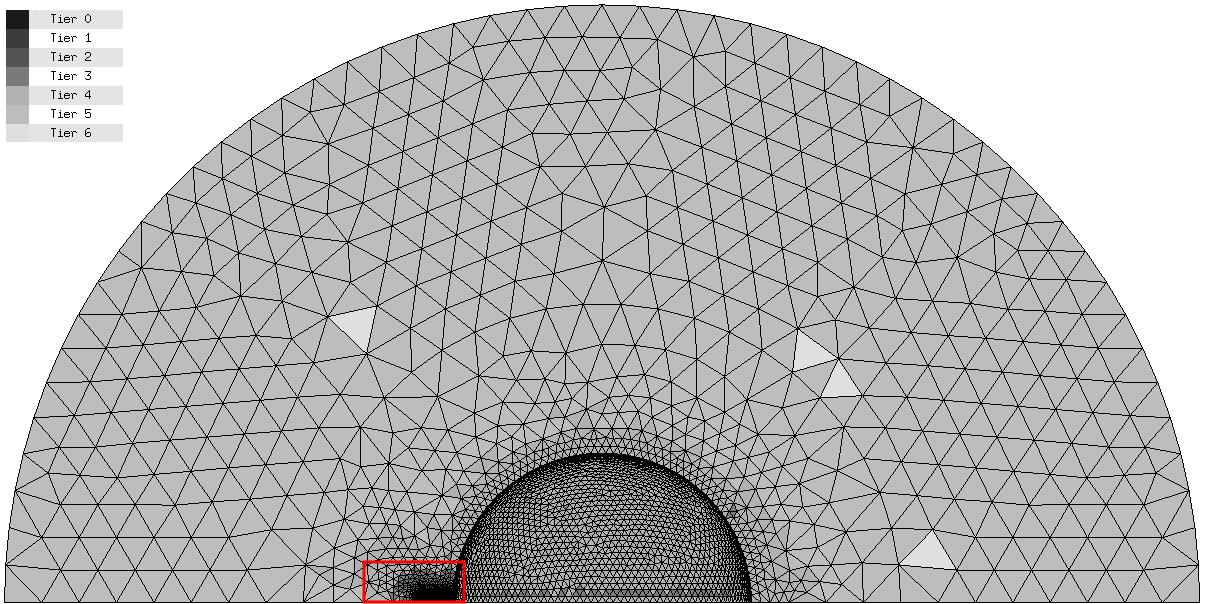}
   \caption{LF2-LTS General view}
  \end{subfigure}    
  \caption{Tier assortment for LF2-LTS}
  \label{fig:antenna-lf2-tierassortment}

 \end{figure}

 In practice, an automated meshing process may produce a quite random tier assortment having an important impact in performance and accuracy.
 This occurs because we let the LTS algorithm and the tier--assortment to span the entire mesh. Notice that in practice this may not be necessary an optimal approach. 
  Figures \ref{fig:rcs-rk-tierassortment}, \ref{fig:rcs-lf2full-tierassortment}, \ref{fig:rcs-lf2-tierassortment}, \ref{fig:antenna-rk-tierassortment}, \ref{fig:antenna-lf2full-tierassortment} and \ref{fig:antenna-lf2-tierassortment} illustrate this phenomenon. 
 For the 1 m PEC sphere (Fig. \ref{fig:rcs-rk-tierassortment}, \ref{fig:rcs-lf2full-tierassortment}, \ref{fig:rcs-lf2-tierassortment}), after imposing a constraint in the element size of $5 \ \text{cm}$ and leaving the rest with $25  \ \text{cm}$ we observe that there is an appreciable amount of scattered elements in the mesh belonging to a lower tier. The meshing algorithm is able to respect the sizes imposed to the elements in the regions closer to the surfaces but not in the inner part. 
 Figures \ref{fig:antenna-rk-tierassortment}, \ref{fig:antenna-lf2full-tierassortment} and \ref{fig:antenna-lf2-tierassortment} represent a variation of the 1m PEC sphere case in which an small cylinder representing a small scale feature has been appended to the sphere. In this example we observe that the presence of scattered lower tiers happens also in problems exhibiting disparate scales, unless the user pre-sets a given maximum number of tiers.
 
  For the LSERK4 algorithm, scattered lower tiers degrade performance because, as depicted in Figure \ref{fig:rcs-rk-stageassortment}, many elements in the neighbourhood of lower tiers have to perform additional operations.
 Additionally, the CPLTS technique requires the storage of the elements in the neighbourhood of smaller tiers, increasing the memory consumption.
  Often the meshing and tier assorting processes result in the highest tier having a very small amount of elements (see Table \ref{table:tierAssortingResAndRCS}), so it is up to the user whether to preserve those tiers or not. In the LF2-CPLTS case, we observe in Figures \ref{fig:rcs-lf2full-tierassortment} and \ref{fig:antenna-lf2full-tierassortment} that the assorting is able to create more tiers than in the LF2-LTS case. This has a positive impact in performance, which is specially relevant in cases with disparate spatial scales such as the presented in Figure \ref{fig:antenna-lf2full-tierassortment}.

 \section{Conclusions} 
 In this work, we have introduced the Causal--Path concept as a way to perform LTS on explicit marching--on--time algorithms. We have applied this concept to the DG discretization under two different time integration techniques: LSERK4 and LF2.
 
 When applied to LSERK4, for all the examples studied, the CPLTS technique has improved the performance by a factor of about two. The dispersive properties of the scheme are not affected while some dissipation is introduced.

 For LF2 the performance is also improved by a factor of about two for a bi--static RCS analysis case. In contrast, the commonly used Montseny's technique provides an speed up of about four. The CPLTS technique however seems to present better dispersive properties than the Montesny's approach and has better adaptivity to multiscale problems.

\bibliographystyle{elsarticle-num}

\section{Bibliograpy}
\bibliography{./library.bib}

\end{document}